\documentclass[manuscript]{acmart}
\AtBeginDocument{%
  \providecommand\BibTeX{{%
    \normalfont B\kern-0.5em{\scshape i\kern-0.25em b}\kern-0.8em\TeX}}}

\setcopyright{acmcopyright}
\copyrightyear{2023}
\acmYear{2023}
\acmDOI{XXXXXXX.XXXXXXX}

\usepackage{amsmath,booktabs, algorithm,natbib}
\usepackage[noend]{algpseudocode}
\usepackage[english]{babel}
\usepackage[utf8]{inputenc}
\usepackage{xcolor}

\newcommand{\change}[1]{{\leavevmode\color{black}{#1}}}

\def\E{\mathbb E}
\def\P{\mathbb P}
\def\R{\mathbb R}

\def\bp{\mathbf p}
\def\br{\mathbf r}

\def\AA{\mathcal A}
\def\BB{\mathcal B}
\def\CC{\mathcal C}
\def\EE{\mathcal E}
\def\HH{\mathcal H}

\def\YY{\mathcal Y}

\begin{document}

\title{Incentive-Aware Recommender Systems in Two-Sided Markets}

\author{Xiaowu Dai}
\affiliation{%
  \institution{University of California,  Los Angeles}
  \country{USA}
}
\email{dai@stat.ucla.edu}

\author{Wenlu Xu}
\affiliation{%
  \institution{University of California,  Los Angeles}
  \country{USA}
  }
\email{wenluxu@g.ucla.edu}

\author{Yuan Qi}
\affiliation{%
  \institution{Fudan University}
  \country{China};
  \institution{Ant Group}
  \country{China}
  }
\email{alanqi0@outlook.com}

\author{Michael I.~Jordan}
\affiliation{%
  \institution{University of California, Berkeley}
  \country{USA};
  \institution{Ant Group}
  \country{China}
  }
\email{jordan@cs.berkeley.edu}

\renewcommand{\shortauthors}{Dai, Xu, Qi, and Jordan}

\begin{abstract}
Online platforms in the Internet Economy commonly incorporate recommender systems that recommend products (or ``arms") to users (or ``agents"). 
A key challenge in this domain arises from myopic agents who are naturally incentivized to \emph{exploit} by choosing the optimal arm based on current information, rather than \emph{exploring} various alternatives to gather information that benefits the collective. 
We propose a novel recommender system that aligns with agents' incentives while achieving asymptotically optimal performance, as measured by regret in repeated interactions. Our framework models this incentive-aware system as a multi-agent bandit problem in two-sided markets, where the interactions of agents and arms are facilitated by recommender systems on online platforms. This model incorporates incentive constraints induced by agents' opportunity costs.
In scenarios where opportunity costs are known to the platform, we show the existence of an incentive-compatible recommendation algorithm. This algorithm pools recommendations between a genuinely good arm and an unknown arm using a randomized and adaptive strategy. 
Moreover,  when these opportunity costs are unknown, we introduce an algorithm that randomly pools recommendations across all arms, utilizing the cumulative loss from each arm as feedback for strategic exploration. We demonstrate that both algorithms satisfy an ex-post fairness criterion, which protects agents from over-exploitation. \change{All code for using the proposed algorithms and reproducing results is made available on GitHub.}
\end{abstract}




\received{NA}
\received[revised]{NA}
\received[accepted]{NA}

\maketitle

\large{\textbf{[2024-06: This paper is to appear at \emph{ACM Transactions on Recommender Systems}.]}}

\section{Introduction}
\label{sec:introduction}

\noindent
Many online platforms in the Internet Economy are organized either explicitly or implicitly as two-sided markets, consisting of products and users, respectively, on the two sides of the online marketplaces.  Moreover, such platforms gather data and use the data to adapt its responses.  In such adaptive markets, the platform acts as a \emph{designer} and plays a dual role: recommending the best product given the information available so far (i.e., exploitation), and trying out less known alternative products to collect more information (i.e., exploration).  
\change{Examples of adaptive online marketplaces include Uber and Lyft, among many others in the Internet Economy.}
The role of exploration is critical to the designer as many products are unappealing ex ante, and few users will find them worthwhile to explore. 
However, exploring these products can be valuable, because the feedback can reveal information about the products and help ascertain whether some of these products are ultimately worthwhile for future users.
But these are marketplaces rather than mere services, and decisions are made by users rather than enforced by the designer. \change{The key challenge arises from the fact that users  may not have an \emph{incentive} to follow the designer's recommendation.}
A myopic user would choose to optimize their reward greedily and has an incentive that skews towards exploitation rather than exploration.
Consequently, the designer may suffer from an insufficient amount of exploration and/or biased data.
For example, if a given product appears worse than users' \emph{opportunity costs} given the information available so far, however noisy the estimate is, this product would remain unexplored even though it may be the best.

In this paper, we study how a designer may incentivize exploration in online marketplaces. 
We consider a designer who can communicate with users; for example,  the designer can send a message and recommend a product to each user, and subsequently observe the user's action and the outcome. 
We focus on a multi-armed bandit (MAB) model with two extra features. First, we consider the multi-agent setting where agents share a common reward distribution on any chosen arm. Second, there is an incentive constraint induced by opportunity costs. 
In this model, each agent (e.g., user) arrives sequentially, and the designer sends a message and recommends an arm (e.g., product) to the agent.   
An agent has a binary action of either following or ignoring the designer's recommendation. The agent receives a reward for a chosen action and immediately leaves the market. However, agents are allowed to return to the market later.
Each agent's action and reward are observed by the designer, but not by other agents. 
An agent has an opportunity cost and would choose an action that satisfies her incentive constraint induced by the opportunity cost.
The designer does not offer payments to agents regardless of agents' actions. 
The designer primarily faces an information design problem of constructing the messages to be sent to agents, and a mechanism design problem of specifying the available arms for each agent that forms the recommendation. 
The information design and recommendation policy altogether influence the actions of myopic agents and incentivize exploration.  The designer's goal is to maximize welfare, which is expressed through regret in repeated games.
\change{Our paper extends the work on incentivized exploration in online marketplaces, building on foundational studies by  \citet{kremer2014implementing, che2018recommender} and \citet{mansour2020bayesian}. We focus on a unique variant of agent incentive constraints and information design in online learning, utilizing historical data to improve recommendations.}

The exact forms of information design and recommendation policy depend on the specific contexts of markets. 
We explore two different realistic scenarios.
The first scenario is when the agents' opportunity costs are \emph{known} to the designer.
From the classical bandit perspective, the optimal policy is to employ an exploration-exploitation tradeoff, such as the upper confidence bound algorithm \citep{lai1985asymptotically} or the active arms elimination algorithm \citep{even2006action}.
However, these algorithms can fail in online marketplaces, because agents are autonomous and may ignore recommendations and refuse to explore if they have no incentive, that is, when their beliefs about the recommended arm are unfavorable. 
For a recommendation policy to be incentive compatible, the agents' beliefs must be favorable toward the recommended arm. We propose an adaptive recommendation policy (ARP) and information design to create such beliefs.
The ARP randomly pools recommendations across a genuinely good arm and an unknown arm. 
The success of this randomized pooling depends on an adaptive explore rate hinging on past agents' actions and rewards. 
It also depends on an information design that keeps agents informed that the exploration is sufficiently infrequent. 
\change{Next, we show that ARP satisfies an ex-post fairness criterion that considers fair treatment across an agent's entire history with the platform. This criterion is specifically designed to ensure that, over time, the proportion of recommendations resulting in unsatisfactory outcomes does not exceed an agent’s tolerance.}  Finally, following the literature on regret minimization, we focus on the asymptotic ex-post regret rate as a function of the time horizon.   We establish that ARP satisfies an ex-post regret rate that is asymptotically optimal.

In the second scenario of interest, each agent's opportunity cost is \emph{private} and \emph{unknown} to the designer. 
Here agents' incentive constraints cannot always be satisfied due to unknown opportunity costs. However, since agents' actions follow their incentive constraints, their actions can be used as feedback for exploration. 
We introduce a modified adaptive recommendation policy (MARP) to incorporate agents' actions for exploration and prevent the optimal arm from being eliminated early: (i) MARP uses each arm's cumulative loss  as feedback that depends on agents' actions. This strategy helps MARP explore arms over time and improves recommendations gradually.
(ii) MARP randomly pools recommendations across all arms. This strategy prevents the optimal arm from being eliminated early. 
Moreover, we show that  MARP also satisfies the ex-post fairness criterion, and establish that MARP can achieve the asymptotically optimal ex-post regret rate. 

Our results have many potential applications in online marketplaces in the Internet Economy. For example, social media such as Instagram recommends content creators based on historical data. Popular creators with content that users found interesting in the past are recommended more often, attracting more visits and reinforcing their popularity. 
In contrast, new creators are often not connected with an audience regardless of the quality of their content.  Our analysis suggests that a randomized recommendation scheme such as ARP and MARP can elevate the visibility of new creators. Moreover, we highlight the incentive constraint: the frequency of randomized recommendations must be kept at a low level so that the audience who receives recommendations about untested creators will find them credible.  Our results quantify the appropriate frequency no matter whether an audience's opportunity cost is known or private.

\subsection{Related Work}
\label{sec:relatedwork}

\noindent
\change{The current paper is rooted in the field of incentivized exploration, initially developed by \citet{kremer2014implementing} and \citet{che2018recommender}.} In the following,  we also provide a broader review of related work from multiple kinds of literature, including mechanism design, matching markets, social learning, and Bayesian persuasion.

\paragraph{Mechanism Design and Incentivized Exploration}
There is a growing body of literature on the intersection of mechanism design and incentivized exploration. When incentives are created via information asymmetry rather than monetary transfers, \citet{kremer2014implementing} derived the Bayesian-optimal policy for a two-arm model. \citet{papanastasiou2018crowdsourcing} considered a similar model but with time-discounted rewards. \citet{bahar2015economic} extended the setting in \citet{kremer2014implementing} to a known social network of agents, where agents observe friends' recommendations but not their rewards. \citet{che2018recommender} analyzed a continuum of agents for a model of two arms with binary rewards.
\citet{mansour2020bayesian} \change{and the follow-ups \citep{mansour2022bayesian, immorlica2020incentivizing, sellke2021price}} extended the BIC policy to a model of multiple arms \change{that allow stochastic bandits as a learning model.
 Note that \citet{mansour2020bayesian} and \citet{immorlica2020incentivizing} considered frequentist regret as a performance measure.
 Our setting is similar to theirs but adopts different incentive constraints that are more suitable for some practical applications and allows agents to return to the platform multiple times. Our adaptive recommendation policy addresses incentivized exploration in online marketplaces while satisfying a fairness criterion.} 
\change{Another closely related line of work considers the ``full revelation'' scenario,  a.k.a. the ``greedy” bandit algorithm, where each agent sees the full history and chooses an action based on her own incentives. Specifically, there are three types of results in this line: positive results for linear contextual bandits under smoothness or
diversity assumptions \citep{kannan2018smoothed, bastani2021mostly, raghavan2023greedy},
positive results for private signals under diversity assumptions \citep{acemoglu2022learning, schmit2018human}, 
and negative results \citep[e.g.,][Ch. 11.2]{slivkins2019introduction}. This corresponds to the full transparency in Section \ref{sec:benchmarkpolicies} and is different from our policy that satisfies the second-best regime under the incentive constraints.}

\paragraph{Matching Markets} \change{The current paper focuses on online marketplaces without monetary payments, which is related to matching markets \citep{gale1962, shapley1971assignment, rothsatomayor1990}.
Many two-sided matching markets function through centralized clearinghouses: medical
residents to hospitals, children to high schools, commissioned officers to military posts, and college
students to dorms.  In principle, centralized clearinghouses act as designers that have the advantage of enforcing agents' decisions and implementing desirable outcomes such as stable matchings. In contrast, our model has a designer that cannot enforce agents' decisions and is thereby different from centralized matching markets. 
There are other matching markets organized in a decentralized manner such that each agent
makes their decision independently of others’ decisions 
\citep{roth1997,dai2020learning,dai2020multi}.   Examples include college admissions, decentralized labor markets, and online dating. In contrast, the  model proposed in this paper allows the designer to coordinate the market through recommendations. Hence it is different from existing models of decentralized matching markets.}  

\paragraph{Social Learning} \change{The field of social learning studies self-interested agents that jointly learn over time in a shared environment  \citep{bikhchandani1992theory, banerjee1992simple, smith2000pathological}. 
Here the agents take actions myopically, ignoring their effects on the learning and welfare of agents in the future.
In particular, \citet{bolton1999strategic} analyzed a continuous-time game of strategic experimentation in multi-agent two-armed bandits, which consists of a safe arm that offers a known payoff and a risky arm of an unknown type. 
\citet{keller2005strategic} studied a similar model but with the feature that the risky arm might yield payoffs after exponentially distributed random times. If the risky arm is good, it generates positive payoffs after exponentially
distributed random times; if it is bad, it never pays out anything. 
\citet{klein2011negatively} and \citet{halac2016optimal} further generalized the strategic experimentation framework to more realistic settings. These models are different from ours as they have no coordination, such as the designer's recommendations.}

\paragraph{Bayesian Persuasion}
\change{The models on Bayesian persuasion consider how a designer can credibly manipulate a single agent's belief and influence her behavior \citep{aumann1995repeated, ostrovsky2010information, rayo2010optimal, kamenica2011bayesian}.  
This is an idealized model for many real-life scenarios in which a more informed designer wishes to persuade the agent to take an action that benefits the designer.  A single round of our model coincides with a version of the Bayesian persuasion game.
Recently, \citet{bergemann2016information} studied information design for messages that the agents receive in a game.
There is also burgeoning literature that studies Bayesian persuasion in dynamic settings \citep{ely2015suspense, halac2016optimal, ely2017beeps}.
In these models, the designer has more information than a single agent due to the feedback from
the previous agents.  Our model also employs such information asymmetry to ensure the desired incentives. The differences are that we consider a multi-agent setting and use the recommendation as a mechanism.}

\subsection{Our Contributions}
\label{sec:contribution}

\noindent
\change{We develop a novel recommender system for online marketplaces that takes into account the agents' incentive constraints.} Our primary methodological and theoretical contributions are summarized as follows.
\begin{itemize}
\item \emph{Incentive-Compatible Algorithms.}
In online marketplaces, classical bandit algorithms are unsuitable due to agents' autonomy and tendency to disregard recommendations if their incentive constraints are not satisfied. To address this challenge, we introduce a novel algorithm called  adaptive recommendation policy (ARP), which builds upon randomization and adaptivity techniques tailored for online marketplaces. \change{The ARP creates positive beliefs among the agents about the recommended arms by strategically pooling recommendations between a genuinely favorable arm and an unknown arm.} Although agents will not knowingly follow a recommendation involving an unknown arm, this pooling of recommendations creates an environment where agents are incentivized to explore. When agents' opportunity cost is known, we prove that ARP guarantees the agents' incentives. 
\item \emph{Extension to Complex Scenarios.} 
We propose a new algorithm, modified adaptive recommendation policy (MARP), designed for complex scenarios where agents' opportunity costs are private and unknown, distinct from ARP in two key respects.
\begin{itemize}
    \item \emph{Feedback Mechanism}. Unlike ARP, which employs the empirical mean of historical rewards to adjust the probability of recommending each arm, MARP leverages each arm's cumulative loss for this purpose. This distinction stems from the fact that, given known opportunity costs, ARP can deduce an optimal arm based on available information up to a specific round, employing it as an exploit arm to satisfy the agent's incentives. \change{Simultaneously, agents can infer the quality of this exploit arm via the empirical mean of historical rewards, thereby determining ARP's maximum exploration rate.} However, with unknown opportunity costs, an agent's action toward a specific arm cannot predict the actions of other agents on the same arm. To address this, MARP tracks the individual loss of each arm and improves recommendations over time.
    \item \emph{Recommendation Pooling}. MARP expands the pooling strategy, randomly pooling recommendations across all arms instead of restricting to two arms as in ARP. This alteration is due to agents possibly having heterogeneous opportunity costs, meaning their reactions to recommendations do not directly reveal the optimal arm for a given round. 
    Consequently, this comprehensive pooling approach prevents prematurely ruling out the optimal arm.
\end{itemize}

\item \change{\emph{Fairness Guarantees.}  We consider the fairness of recommendation systems in online marketplaces where agents frequently interact with the platform. We propose an ex-post fairness criterion that considers fair treatment across an agent's entire history with the platform. This criterion is specifically designed to ensure that, over time, the proportion of recommendations resulting in unsatisfactory outcomes does not exceed an agent's tolerance. We prove that ARP and MARP satisfy this ex-post fairness requirement, making them different from existing recommendation policies such as the algorithms in \citet{mansour2020bayesian}, which fail to guarantee this form of fairness uniformly across all sequences of agents. 
Moreover, the ex-post fairness differs from existing fairness criteria, such as the fair exploration concept in \citet{kannan2017fairness}, which ensures worse arms will not be preferred to better ones. In contrast, ex-post fairness shifts the focus to the opposite side of the market, providing safeguards to ensure that returning agents are not subjected to over-exploitation. This new perspective aligns the marketplace dynamics more closely with practices.}
\item \emph{Sublinear Regrets.} 
ARP achieves an asymptotically optimal welfare, characterized by the cumulative regret across all agents. 
To state this more rigorously, consider $T$ as the time horizon and $m$ as the number of arms. Then ARP yields an upper bound for the ex-post regret given by,
\begin{equation*}
L^*+O\left(\sqrt{mT\ln(mT)}\right),
\end{equation*}
where $L^*$ is a constant, and this regret bound is sublinear with respect to the  time horizon $T$. Moreover, MARP also achieves a sublinear regret. 
\end{itemize}
\noindent
The rest of this paper is organized as follows. 
Section \ref{sec:model} introduces the model of recommendation in online marketplaces under an incentive constraint. 
Section \ref{sec:learnreveal} presents the algorithm ARP, showing that it is incentive compatible and fair under known opportunity costs.
Section \ref{sec:benchmarkpolicy} studies three benchmark policies under known opportunity costs.
Section \ref{sec:privatereveal} presents the algorithm MARP that minimizes the regret and is fair under private opportunity costs.
Section \ref{sec:relatedwork} discusses related work.
Section \ref{sec:discussion} concludes the paper with further research directions.  All proofs are provided in the Appendix.

\section{Model}
\label{sec:model}

\noindent
This section introduces a recommendation model with incentive constraints for online marketplaces. Following the multi-armed bandit (MAB) terminology, we respectively call participants on each side as \emph{agents} (e.g., users of the platform) and \emph{arms} (e.g., products).
This work generalizes the MAB setting in several directions, both in terms of the information design problem which consists in the design of messages to agents and in terms of the mechanism design problem being solved by the designer's recommendation.
We contrast the mechanism design problem with the information design problem, where a mechanism specifies the available actions for each agent, but does not control the information structure \citep{bergemann2016information}.

\subsection{Interaction Protocol}
 
\noindent
Consider an online marketplace in which a sequence of $T$ agents arrives sequentially and the \change{\emph{designer} (e.g., the platform)} recommends an arm to an agent. 
The interaction protocol between agents and designer is as follows. At each round $t\in[T]\equiv\{1,\ldots, T\}$, agent $t$ arrives and observes a message $\EE_t$ sent by the designer, where $\EE_t$ is a subset of the information that the designer has collected from all past agents. Next, the designer recommends an arm $I_t\in\AA\equiv\{1,\ldots,m\}$ to agent $t$.
The recommendation is private, and each agent only observes the recommendation made to her.
Agent $t$ chooses an action $y_t\in\YY\equiv\{0,1\}$, where $y_t=1$ if she \emph{pulls} the recommended arm, and $y_t=0$ if she \emph{ignores} recommendation and does not pull any arm. Finally, agent $t$ leaves the market before round $t+1$. 
\change{We illustrate this interaction protocol on two real-world platforms. First, on Instagram, upon visiting the platform, users are presented with recommended content, which might appear as a highlighted video in the user's feed or as a notification. In this context, pulling the arm means that users watch the recommended video. Choosing to ignore the recommendation corresponds to viewing the notification but not engaging with the content.
Second, in the case of Uber, when a customer uses the service, the platform sends a message recommending rides based on the customer’s request. Here, pulling the arm means accepting one of these recommendations and proceeding with a ride, while ignoring the recommendation means that the customer evaluates the recommended rides but decides not to take the provided options.}

\begin{figure}[ht!]
\centering
\includegraphics[width=0.5\textwidth]{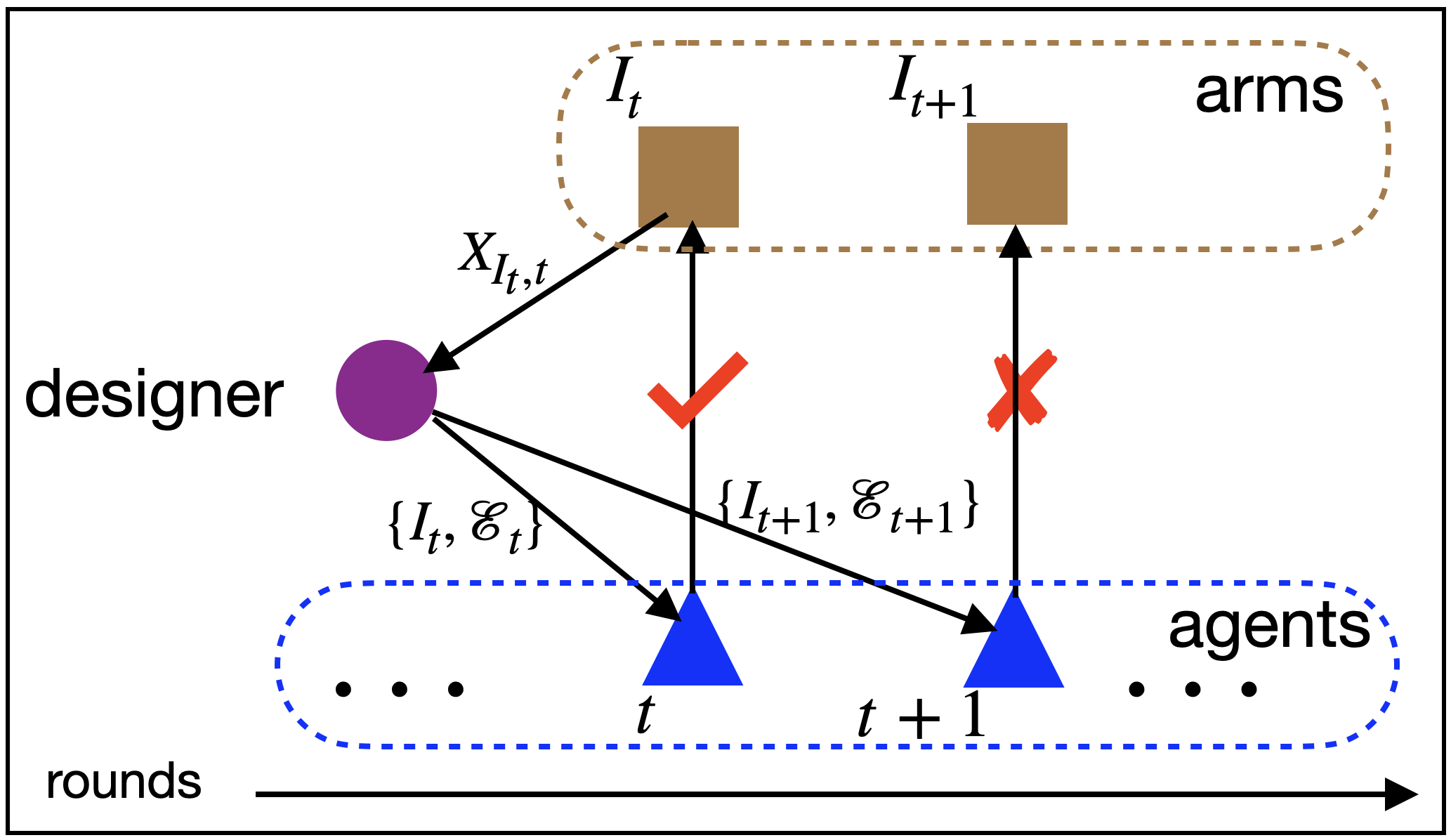}
\caption{\change{The recommendation process mediated by the designer for online marketplaces.}}
\label{fig:flow}
\end{figure}

Figure \ref{fig:flow}  provides an illustration of the process.
This model of interaction protocol is equivalent to classical MAB  if each agent always pulls the recommended arm. In other words, two models are equivalent if $y_t= 1$ for any $t\in[T]$. Then a recommendation algorithm would become a bandit algorithm.
\subsection{Agent's Incentive}
 
\noindent
When an agent $t$ pulls an arm $i\in\AA$, the agent receives a stochastic reward $X_{i,t}$ that is drawn i.i.d.\ from $[0,1]$ with  mean $\mu_{i}$.
\change{For a recommendation policy to be implementable, agents must be incentivized to follow the recommendations. Since the specific context of the recommendation (whether for exploration or exploitation) is kept hidden from the agents, their incentive to follow it depends on their perception of the designer’s information.}
We consider that all agents are \emph{myopic} and choose actions that satisfy an \emph{incentive constraint} specified as follows. Each agent $t$ has an \emph{opportunity cost} $c_t\in(0,1)$ for pulling an arm, where $c_t$ can be the time spent or the price charged. 
Conditional on $I_t$ and $\EE_{t}$, we define the agent $t$'s incentive constraint as,
\begin{equation}
\label{eqn:incentivesct}
\E[X_{I_t,t}|I_t=i,\EE_{t}]\geq c_t,\quad\forall t\in[T],\ \forall i\in\AA.
\end{equation}
Agent $t$ chooses $y_t=1$ if and only if Eq.~\eqref{eqn:incentivesct} holds.
\change{A similar incentive constraint has been used by \citet{che2018recommender}, which studied the two-arm case, $m=2$. Our model in Eq.~\eqref{eqn:incentivesct} extends to any set $\mathcal{A}$ with $m \geq 2$.}
Moreover, Eq.~\eqref{eqn:incentivesct} implies that $y_t$ is a random variable that depends on information $\EE_{t}$ and recommendations $\{I_1,\ldots,I_{t}\}$.

\change{The incentive constraint in Eq.~\eqref{eqn:incentivesct}  differs fundamentally from  Bayesian Incentive Compatibility \citep[BIC, ][]{kremer2014implementing, papanastasiou2018crowdsourcing, mansour2020bayesian}. BIC requires each agent’s Bayesian expected reward to be maximized by the recommended action, conditional on the information flow controlled by the designer, and each agent can evaluate all arms.} 
\change{In contrast, Eq.~\eqref{eqn:incentivesct} requires the recommended arm to be at least as good as the opportunity cost. This approach is particularly useful in large marketplaces, like ride-sharing platforms, where the large number of similar drivers prevents users from evaluating all available options due to time constraints, and where platforms restrict information by only showing recommended options to each user, as exemplified by Uber, which presents only recommended drivers to its users.}

\subsection{Information Design for Recommendation}
\label{sec:infodesign} 

\noindent 
Providing future agents with access to historical data is a key value proposition in many online marketplaces. The designer faces an information design problem of constructing the message 
\begin{equation*}
\EE_t\subseteq \left\{(I_s,y_s,X_{I_s,s}): 1\leq s\leq t-1\right\}.
\end{equation*} 
The design of $\EE_t$ needs to consider constraints arising from practical and legal issues that are typically set against information asymmetry. 

\change{We are interested in \emph{transparent} mechanisms for $\EE_t$, which disclose the reward history without revealing the identities of recommended arms. Consequently, $\mathcal E_t$ does not depend on the arm identities $I_s$, but only on the reward history $(y_s,X_{I_s,s})$ for $1\leq s\leq t-1$. Thus, under transparent mechanism, we have,
\begin{equation*}
\mathcal E_t\subseteq  \left\{(y_s,X_{I_s,s}): 1\leq s\leq t-1\right\}.
\end{equation*}}
For example, the design of $\EE_t$  for a specific restaurant would disclose the customers' historical scores and reviews about the restaurant while protecting the identities of specific dishes chosen by each customer.  
The explicit forms of $\EE_t$ depend on whether each agent's opportunity cost $c_t$ in Eq.~\eqref{eqn:incentivesct} is known or private to the designer, which will be discussed in a later section.
\change{We note that for the legal implications of using historical data for making recommendations, under data protection regulations such as the General Data Protection Regulation (GDPR) in the European Union \citep{hoofnagle2019european}, our approach is compliant as long as it involves transparent data handling processes.}

\change{Our model operates within an online-learning environment. A key feature of our approach is to use the  observed historical rewards, together with information design for recommendation. We do not need the prior assumptions on the mean rewards $\mu_i$ for $i\in\mathcal A$.  In contrast, standard Bayesian models \citep[e.g.,][]{mansour2020bayesian},  typically focus on analyzing incentives based on random rewards and their distributions. These models generally require the assumption of priors on the mean rewards $\mu_i$ for $i\in\mathcal A$ as a crucial part of their analysis, particularly for satisfying the incentive constraint.}

\subsection{Mechanism Design for Recommendation}
\label{sec:mechdesignandrecom} 
We also face the mechanism design problem of specifying the available arms for each agent. 
Different from information design, the mechanism design problem does not control the information structure and is solved by the designer's recommendation.
We focus on a \emph{randomized recommendation} scheme in which the designer chooses a vector $\bp_t=(p_{1,t},\ldots,p_{m,t})$ over the set $\AA$ and recommends arm $j$ with a probability $p_{j,t}\geq 0$, where $p_{1,t}+\cdots+p_{m,t}=1$ and $t\geq 1$. 
\change{To implement the randomized recommendation, let $U_1, U_2, \ldots$ be the observed outcomes of random variables that are independently and uniformly drawn from $[0, 1)$}. Then the  designer's recommendation $I_t$ at round $t$ is chosen by,
\begin{equation}
\label{eqn:randrecpolicy}
\begin{aligned}
 I_t =  \sum_{j\in\AA} j\cdot \mathbf{1}\left\{U_t\in\left[\sum_{s=1}^{j-1}p_{s,t},\sum_{s=1}^jp_{s,t}\right)\right\}.
 \end{aligned}
\end{equation}
\change{That is, the arm $I_t$ is chosen by finding the smallest index $j$ such that the cumulative sum of probabilities up to $p_{j,t}$ exceeds $U_t$. This mechanism ensures that each arm is chosen with the probability specified by $\mathbf{p}_t$.}
The probability $\bp_t$ is determined entirely by agents' past actions and rewards, and does \emph{not} depend on the past recommendations $I_1, \ldots, I_{t-1}$.
The explicit form of  $\bp_t$ depends on whether the opportunity cost $c_t$ in Eq.~\eqref{eqn:incentivesct} is known or private to the designer, which will be discussed later in this paper.

\change{In contrast to the policy in Eq.~\eqref{eqn:randrecpolicy},  \citet{mansour2020bayesian} propose a different type of recommendation strategy. Their approach fixes a pool of future agents a priori  and recommends an explore arm to a fraction within this pool. Consequently, every agent in this pool faces the same probability of exploration. The Eq.~\eqref{eqn:randrecpolicy}, however, offers an individualized exploration probability, where $p_{j,t}$'s can vary for each agent. This flexibility allows us to integrate a fairness metric into the recommendation policy, thereby enhancing the quality of the recommendations. }

\subsection{Designer's Objective}

\noindent
The designer receives a signal about the recommended arm $I_t$ in the form of agent $t$'s action $y_t\in\YY$, and the reward $X_{I_t,t}$ if $y_t=1$. 
We consider  the following loss for the designer,
\begin{equation}
\label{eqn:loss}
l(I_t,y_t) = \max_{i\in\AA}\E[X_{i,t}  - X_{I_t,t}\cdot y_t],
\end{equation}
where the expectation is taken over randomized recommendations and stochastic rewards. Thus, $l(\cdot,\cdot)$ measures the loss for competing against the arm that is optimal in expectation, which excludes the random fluctuations in stochastic rewards. Moreover,  this loss function depends on the agent's action $y_t$ and can be nonconvex in its first argument $I_t$.
The  designer's objective is to maximize social welfare. Equivalently, the designer aims to minimize the \emph{regret} defined by,
\begin{equation}
\label{eqn:defregret}
R_T = \sum_{t=1}^Tl(I_t,y_t) - \min_{i\in\AA}\sum_{t=1}^Tl(i,y_t).
\end{equation}
Here $R_T$ measures the realized difference between the cumulative loss and the loss of the optimal arm.
Compared to the classical MAB, which also aims to minimize regret, a significant challenge in our setting is that the action $y_t$ is subject to the additional incentive constraint in Eq.~\eqref{eqn:incentivesct}.

\change{Besides the social welfare, we also consider the objective of \emph{fairness} in online marketplaces. Since agents may repeatedly engage with the market, an important issue is whether agents are treated fairly over time. We introduce an ex-post fairness criterion that accounts for an agent's history of interactions. 
Consider that each agent $t$ has a context defined by $(\alpha_t,\beta_t)$, where $\alpha_t$ is the total number of agent $t$'s visits up to (but not including) round $t$, and $\beta_t$ represents the number of those visits where the agent followed the recommendation but received a reward worse than their opportunity cost $c_t$.
Let $\gamma_t\in[0,1]$ be the agent $t$'s tolerance level for exploration that is known to the designer. 
A policy is considered \emph{ex-post fair} if it ensures that the proportion of unsatisfactory recommendations does not exceed the agent's tolerance level,
\begin{equation}
\label{def:expostfair}
    \frac{\beta_t+\mathbf{1}\{\E[X_{I_t,t}]<c_t\}}{\alpha_t+1}\leq \gamma_t,\quad\forall t\geq 1.
\end{equation}
This criterion protects agents, both new and returning, from over-exploitation. 
For new agents $(\alpha_t=\beta_t=0)$,  the designer recommends exploitation arms unless $\gamma_t=1$.
For returning agents with historical data $(\alpha_t,\beta_t)\neq(0,0)$,  the unsatisfactory rate is controlled by $\gamma_t$. 
Our definition of ex-post fairness differs from existing fairness criteria, such as the ``fair exploration" concept by \citet{kannan2017fairness}, which focuses on not preferring worse arms over better ones. In contrast, ex-post fairness emphasizes protecting agents from over-exploitation, offering a new perspective that aligns the marketplace dynamics with ethical practices.
}

\section{Optimal Policy Under Known Opportunity Costs}
\label{sec:learnreveal}

\noindent
We start with the scenario where agents' opportunity costs are known to the designer.
To fix ideas, assume the costs are homogeneous, i.e., $c_t= c_*$ for $t\geq 1$. Then the incentive condition in Eq.~\eqref{eqn:incentivesct} is equivalent to
\begin{equation*}
\E[X_{I_t,t}|I_t=i,\EE_{t}]\geq c_*, \quad\forall t\in[T] \text{ and } i\in\AA.
\end{equation*}
Assume that arm $i=1$'s expected reward is known and greater than the cost,
\begin{equation}
\label{eqn:positiveplat}
\mu_1>c_*.
\end{equation}
\change{This assumption is necessary because the problem becomes intractable if the expected rewards of all arms are unknown or if all expected rewards are less than the cost.  A similar assumption is also made in Property 1 by \citet{mansour2020bayesian}.
In online marketplaces, initial knowledge of an arm's expected reward, such as $\mu_1$, might arise from product research. For example, Pandora studied the detailed attributes of music through their music genome project \citep{joyce2006pandora}.  It might also arise from a  flow of users who face negligible exploration costs and, therefore, do not mind exploring the arm. Except for arm $i=1$, we assume that other arms, $i=2, \ldots, m$, have a priori unknown expected rewards. A fundamental challenge for the designer is to design  a mechanism for sending the message $\mathcal{E}_t$, which will incentivize agents to explore other arms. In this section, we propose a new algorithm that provides the appropriate incentive and satisfies  the fairness criterion.}

\subsection{Adaptive Recommendation Policy}
\label{sec:ARP}
\noindent
We present an algorithm called \emph{adaptive recommendation policy} (ARP), built upon concepts of ``randomized recommendation" and ``adaptive exploration.” Specifically, ARP employs the randomized recommendation in Eq.~\eqref{eqn:randrecpolicy}, where it selects an arm $i$ according to a probability distribution $\{p_{i,t}:i\in\AA\}$ and then recommends it to an arriving agent $t$. Moreover, the probability distribution is adaptive to agents' historical rewards. If an arm yields inferior rewards, the exploration rate of the subsequent arm will decrease. This adaptivity, in turn, incentivizes the agents' exploration.

\begin{algorithm}[ht]
\caption{ \normalsize{{Adaptive Recommendation Policy (ARP)}}}
\label{alg:adaptivesampling}  
\begin{algorithmic}[1]
\State  \normalsize{\textbf{Step 1 (Input):} Set $\AA$; Parameters: $\lambda, \theta_\tau, k>0$ in Eqs.~\eqref{eqn:choiceoflambda}, \eqref{eqn:thetatau} and \eqref{eqn:choiceofk}, respectively.}
\State  \normalsize{\textbf{Step 2 (\change{Sampling}):}}  Announce $\EE_t$ according to Eq.~\eqref{eqn:publiceetinitial}. Recommend arm $i=1$ to the first $k$ agents and collect rewards $X_{1,1},\ldots,X_{1,k}$. Let $\hat{\mu}_1^k=\frac{1}{k}\sum_{s=1}^kX_{1,s}$. 
\State \textbf{for} each arm $i>1$ in increasing lexicographic order \textbf{do}
\State \quad\quad Choose $a_i^*$ and $\bp_{t}$  by Eqs.~\eqref{eqn:choiceofai*} and \eqref{eqn:choiceofLi}; Update $\EE_t$ by Eq.~\eqref{eqn:publiceet}; Set $\tau_i=0$, $L_i=0$;
\State \quad\quad \textbf{while} the counter $\tau_i<k$ \textbf{do}
\State \quad\quad\quad\quad Recommend $I_t$ by Eq.~\eqref{eqn:randrecpolicyindfair}; 
\State \quad\quad\quad\quad \quad\quad If $I_t=i$, update $\tau_i=\tau_i+1$;
\State \quad\quad\quad\quad Set $t=t+1$, $L_i=L_i+1$;
\State \quad\quad \textbf{end}
\State \quad\quad Collect $k$ rewards of arm $i$ and calculate the mean reward $\hat{\mu}_i^k$.
\State \textbf{end}
\State \normalsize{\textbf{Step 3 (Exploration):}}  Update $\EE_t$ according to Eq.~\eqref{eqn:publiceetall}. Initialize $\BB=\AA$,  $q=k$. 
\State \textbf{while} the arm set size $|\BB|>1$ \textbf{do}
\State \quad\quad Let $\hat{\mu}_i^q$ be the mean reward of arm $i$ based on $q$ rewards;
\State \quad\quad  Let $\hat{\mu}^q_* = \max_{i\in\BB}\hat{\mu}_i^q$, and $\BB = \left\{i:\hat{\mu}_i^q+\sqrt{\frac{\ln(T\theta_\tau)}{2q}}\geq \max\{\hat{\mu}_*^q, c_*\}\right\}$;
\State \quad\quad For the next $|\BB|$ agents, recommend each arm $i\in\BB$;  Set $q=q+1$.
\State \textbf{end}
\State \normalsize{\textbf{Step 4 (Exploitation):}} For the remaining agents, recommend the arm left in $\BB$.
\end{algorithmic}
\end{algorithm}

We summarize ARP in Algorithm \ref{alg:adaptivesampling}, and now detail the four main steps.
The first step is the input of key parameters. Let $\lambda>0$ be a margin parameter that satisfies,
\begin{equation}
\label{eqn:choiceoflambda}
0<\lambda\leq \frac{3}{4}(\mu_1-c_*).
\end{equation}
Then the upper bound of $\lambda$ depends on the gap between $\mu_1$ and $c_*$. 
Let $\tau\in(0,1-c_*)$, and
\begin{equation}
\label{eqn:thetatau}
\theta_\tau = \frac{4m^2}{\tau\cdot\min_{i\in\AA}\P(\mu_i-c_*\geq \tau)}.
\end{equation}
For example, if the prior distribution of $\mu_i$ is uniform on $[0,1]$, then $\min_{i\in\AA}\P(\mu_i-c_*\geq \tau) = 1-c_*-\tau>0$.
Given $\theta_\tau$, the designer selects a sample size $k>0$ satisfying,
\begin{equation}
\label{eqn:choiceofk}
k\geq \max\left\{\frac{9}{2\lambda^2}\ln\left(\frac{20m}{\lambda}\right),\ \theta_\tau^2\ln(T\theta_\tau)\right\},
\end{equation}
where $m$ is the number of arms, and $T$ is the time horizon.
\change{The margin parameter $\lambda > 0$, as defined in Eq.\eqref{eqn:choiceoflambda}, ensures that when the sample size $k$ is sufficiently large to satisfy Eq.\eqref{eqn:choiceofk}, there is a high probability that the expected reward of the recommended arm will exceed the opportunity cost. This is proven in Section \ref{sec:agentincentive}.}

The second step is \change{the sampling of rewards for all arms}. 
We split this step into  $m$ stages, where each stage is indexed by $i=1,\ldots,m$. 
The goal is to collect $k$ samples for each arm $i$.
Let $L_i$ be the number of rounds at each stage $i$, which is specified later.
Consider the initial stage $i=1$. 
The designer sends the message $\EE_t $  to agent $t\in\{1,\ldots,k\}$, where
\begin{equation}
\label{eqn:publiceetinitial}
\EE_t = \left\{\mu_1>c_*\right\}\cup \left\{\text{exploration rate at time $t$ is }0\right\}.
\end{equation}
Then designer recommends arm $i=1$ to agents $t\in\{1,\ldots,k\}$ according to Eq.~\eqref{eqn:randrecpolicy}. Here $\bp_t$   is given by $p_{1,t} = 1$ and $p_{j,t}=0$ for any  $j\neq 1$.
Conditional on $\EE_t$ in Eq.~\eqref{eqn:publiceetinitial}, each agent $t$ is aware that the recommended arm has a larger expected reward than the opportunity cost. 
Then the designer collects rewards $X_{1,1},\ldots,X_{1,k}$, and calculates the empirical mean reward $\hat{\mu}_1^k = \frac{1}{k}\sum_{s=1}^kX_{1,s}$ of arm $i=1$.  The initial stage has a total of $L_1=k$ rounds.
Next, we consider stage $i>1$. Here the designer chooses arm $i$ as the explore arm, and it chooses an exploit arm $a_i^*$ by
\begin{equation}
\label{eqn:choiceofai*}
a_i^*=\underset{1\leq j\leq i-1}{\arg\max}~\hat{\mu}^k_j,
\end{equation} 
where ties are broken randomly. The designer specifies the vector $\bp_t$ in  Eq.~\eqref{eqn:randrecpolicy}  as, 
\begin{equation}
\label{eqn:choiceofLi}
p_{i,t} = \frac{\lambda}{2(c_*-\widehat{M}_i)+\lambda}\mathbf{1}\left\{\widehat{M}_i< c_*\right\} + \mathbf{1}\left\{\widehat{M}_i\geq c_*\right\},  \quad p_{a_i^*,t} = 1-p_{i,t},  
\end{equation}
and $p_{j,t} =0, \forall j\not\in\{ a_i^*,i \}$. Here $\sum_{j=1}^{i-1}L_{j}+1\leq t\leq \sum_{j=1}^iL_j$, and $\widehat{M}_i$ in Eq.~\eqref{eqn:choiceofLi} is the empirical mean of historical rewards up to the stage $i-1$,
\begin{equation}
\label{eqn:hatmi}
\widehat{M}_i = \frac{1}{\sum_{j=1}^{i-1}L_j}\sum_{s=1}^{L_1+\cdots+L_{i-1}}X_{I_s,s} \cdot y_s.
\end{equation} 
It is seen that the exploration rate $p_{i,t}$ is adaptive to the historical rewards through $\widehat{M}_i$. 
Moreover, the message $\EE_t$ in stage $i>1$ is updated as follows,
\begin{equation}
\label{eqn:publiceet}
\EE_t = \left\{\mu_1>c_*\right\}\cup \left\{\text{exploration rate }p_{i,t}\right\}\cup \left\{X_{I_s,s}: y_s=1 \text{ and } 1\leq s\leq\sum_{j=1}^{i-1}L_j\right\}.
\end{equation}
Then under $\EE_t$ in Eq.~\eqref{eqn:publiceet}, agent $t$ does not observe past agents' actions or identities of arms that were recommended to past agents. Indeed, $\EE_t$ only discloses the information that $\AA$ includes a good arm, together with the exploration rate $p_{i,t}$ at the current stage, and the reward history of all agents who have followed the recommendations.  
\change{Agents can form a rational belief about the designer's beliefs regarding the recommended arms. The probability of the arm $I_t$ being recommended for exploration is $p_{i,t}$, which is a part of the message $\mathcal E_t$ sent to the agent $t$, and the probability of the arm being recommended for exploitation is $1-p_{i,t}$. The belief formation allows agents to adjust their incentives accordingly at each time $t$.} Finally, the designer can incorporate the fairness constraint into the recommendation. Suppose that the designer collects historical data $(\alpha_t,\beta_t)$ for agent $t$, and forms a criterion $\HH(t)$ where agent $t$ would like to return even if the agent receives an unsatisfactory recommendation,
\begin{equation}
\label{eqn:egofht}
\HH(t) = \left\{\frac{\beta_t+1}{\alpha_t+1}\leq\gamma_{t}\right\},\quad\forall t\geq 1.
\end{equation}
Note that $\HH(t)$ applies to the setting where agents do not return to the market because we can simply set $\gamma_t=1$. 
The designer  can employ  the  policy,
\begin{equation}
\label{eqn:randrecpolicyindfair}
\begin{aligned}
 I_t =   \sum_{j\in\{a_i^*,i\}} j\cdot \mathbf{1}\left\{U_t\in\left[\sum_{s=1}^{j-1}p_{s,t},\sum_{s=1}^jp_{s,t}\right)\right\}\cdot \mathbf{1}\{\HH(t) \text{ holds}\} + a_i^*\cdot\mathbf{1}\{\HH(t) \text{ fails}\}.
 \end{aligned}
\end{equation}
Here $p_{s,t}$ is given by Eq.~\eqref{eqn:choiceofLi} and $\sum_{j=1}^{i-1}L_{j}+1\leq t\leq\sum_{j=1}^iL_j$. 
This policy is a generalization of the randomized recommendation in Eq.~\eqref{eqn:randrecpolicy}, by imposing the  criterion $\HH(t)$.
The stage $i$ is concluded after the designer collects a total of $k$ rewards of arm $i$, and the designer can calculate the number of rounds $L_i$ and arm $i$'s mean reward $\hat{\mu}_i^k$  based on $k$ rewards.

The third step is the \emph{exploration} of arms. After collecting $k$ rewards of each arm in $\AA$, the designer updates $\EE_t$ as follows,
\begin{equation}
\label{eqn:publiceetall}
\EE_t = \left\{k \text{ rewards that were collected for each arm } i\in\AA\right\}.
\end{equation}
Next, arms are compared to one another and compete against the cost $c_*$. This step is an incentivized version of the Active Arms Elimination algorithm in \cite{even2006action}. Specifically, each arm is initially labeled \emph{active} and included in a set $\BB$. 
Let  $\hat{\mu}_i^q$ be the empirical mean reward of arm $i$ based on $q$ rewards,  where $q\geq k$.  Let $\hat{\mu}^q_* = \max_{i\in\BB}\hat{\mu}_i^q$. Define an arm $i$ to be suboptimal if the following condition holds,
\begin{equation*}
\hat{\mu}_i^q+\sqrt{\frac{\ln(T\theta_\tau)}{2q}}<\max\left\{\hat{\mu}_*^q, c_*\right\}.
\end{equation*}
Here the designer compares each arm with the cost $c_*$ and guarantees the agent's incentive constraint.
The suboptimal arms will be permanently eliminated from $\BB$.
For the next $|\BB|$ agents, the designer recommends each arm $i\in\BB$ according to an increasing lexicographic order and collects a new reward for each arm $i$.  The algorithm repeats the comparison and drops arms over time until only one arm remains.

The fourth step is the \emph{exploitation} of an arm, where the designer commits to recommending the single arm in $\BB$ to the remaining agents. \change{We note that there are structural similarities between  the ARP in Algorithm \ref{alg:adaptivesampling} and those algorithms in \citet{mansour2020bayesian} and \citet{sellke2021price}, particularly in the shared structure of initial input, followed by stages of sampling, exploration, and exploitation.  However, a unique feature of ARP is its use of randomized recommendations within online recommendation systems. }

\subsection{Incentive Guarantee} 
\label{sec:agentincentive}
\noindent
We now study the performance of ARP in Algorithm \ref{alg:adaptivesampling}. First, we consider the incentive guarantee.
The exploit arm $a_i^*$ in  Eq.~\eqref{eqn:choiceofai*} is based on empirical mean rewards, whereas the incentive constraint in Eq.~\eqref{eqn:incentivesct} is defined based on the expected rewards.
Although there exists a discrepancy between the empirical and expected rewards, we show in Theorem \ref{thm:knowncost} that ARP guarantees the agent's incentive.
The key is that we quantify a margin parameter $\lambda>0$ in Eq.~\eqref{eqn:choiceoflambda} to ensure that when the sample size $k$ is large enough to satisfy  Eq.~\eqref{eqn:choiceofk}, there is a high probability that the expected reward of the recommended arm is greater than the opportunity cost. Indeed, if $\lambda=0$, then $k$ needs to be infinitely large to guarantee the agent's incentive.
\begin{theorem}
\label{thm:knowncost}
Suppose that the assumption in Eq.~\eqref{eqn:positiveplat} holds and  parameters $\lambda,\theta_\tau, k$ are chosen according to Eqs.\ \eqref{eqn:choiceoflambda}, \eqref{eqn:thetatau}, and \eqref{eqn:choiceofk}, respectively.
Then ARP in Algorithm \ref{alg:adaptivesampling} guarantees the agent's incentive in Eq.~\eqref{eqn:incentivesct} for any $t\in[T]$ and $i\in\AA$.
\end{theorem}
\noindent
The ARP has an exploration rate $p_{i,t}$  that is adaptive to historical rewards. Specifically,
if $\widehat{M}_i<c_*$ and the explore arm $i$ yields inferior rewards with $\widehat{M}_{i+1}\leq\widehat{M}_i$, then  the exploration rate in  Eq.~\eqref{eqn:choiceofLi}  for arm $i+1$ would decrease with $p_{i+1,t}\leq p_{i,t}$. If $\widehat{M}_i\geq c_*$,  then the exploration rate in  Eq.~\eqref{eqn:choiceofLi}  can take its maximum value as $p_{i,t}=1$. By Theorem \ref{thm:knowncost}, this adaptivity is sufficient to guarantee the agent's incentive. On the other hand, this adaptivity also accelerates exploration. 
Consider a policy with a fixed exploration rate, then we need to set the rate as $\min_{1< i\leq m}\min_{L_1<t\leq T}p_{i,t}$  for each stage $i>1$ in order to guarantee the agent's incentive. However, this fixed exploration rate may be significantly smaller than most of $p_{i,t}$'s with $i>1$ and $t>L_1$,  and hence makes the exploration much slower.

\subsection{Ex-Post Fairness}
\label{sec:indfairness}
\noindent
The randomized policy ARP incorporates the constraint $\HH(t)$ defined in Eq.~\eqref{eqn:egofht} for any $t\geq 1$. We demonstrate that ARP meets the criterion for ex-post fairness.
\begin{theorem}
\label{lem:individualfair}
The ARP  procedure in Algorithm \ref{alg:adaptivesampling} satisfies the ex-post fairness criterion in Eq.~\eqref{def:expostfair}.
\end{theorem}

\change{We compare ARP with the policy in \citet{mansour2020bayesian}, identifying four key differences. First, the policy in Mansour et al.\ cannot uniformly guarantee ex-post fairness across all agent sequences. This policy assigns a fixed pool of agents to explore each arm, giving every agent in the pool an equal probability of exploration. However, this approach can result in a sequence of agents where historical data do not satisfy the criterion $\HH(t)$ for some instances $t$. Second, the incentive structure of ARP and Mansour et al.\ differ significantly. ARP is designed to satisfy the incentive condition in Eq.~\eqref{eqn:incentivesct}, whereas Mansour et al.\ follow a Bayesian incentive compatibility (BIC) framework. Third, Mansour et al.\ policy uses a exploration rate  that is not adaptive to agent's historical rewards. Following the discussion of Theorem \ref{thm:knowncost}, this nonadaptive policy leads to a slower exploration compared to the adaptive policy of ARP, which dynamically adjusts exploration rate based on agent's historical data.
Fourth, there is a contrast in information transparency between the two policies. Mansour et al.\ require agents to know the identity of the recommended arm, but not its reward history. In contrast, ARP shares arms' reward histories while keeping their identities undisclosed.}

\subsection{Regret Analysis}
\noindent
We provide an upper bound of the regret for ARP.  Note that the definition of  the regret $R_T$ in Eq.~\eqref{eqn:defregret}  depends on the agent's action $y_t$. However, since Theorem \ref{thm:knowncost} respects the agent's incentive constraint, we show that ARP can achieve the optimal regret bound as if the regret $R_T$ does not depend on the agent's action. 
\begin{theorem}
\label{lem:regretofknowncost}
Suppose the assumption in Eq.~\eqref{eqn:positiveplat} holds, and the parameters $\lambda,\theta_\tau, k$ are chosen by Eqs.~ \eqref{eqn:choiceoflambda}, \eqref{eqn:thetatau}, and \eqref{eqn:choiceofk}, respectively.
Let $\gamma_t=1$ in Eq.~\eqref{eqn:egofht}. Then the ARP procedure in Algorithm \ref{alg:adaptivesampling} achieves ex-post regret,
\begin{equation*}
\begin{aligned}
R_T &\leq L^*+O\left(\sqrt{mT\ln(mT)}\right),
\end{aligned}
\end{equation*}
where $L^* \equiv L_1+\cdots+L_m$ is the total number of rounds in Step 2 of ARP, $m$ is the number of arms, and $T$ is the time horizon.
\end{theorem}

\section{Benchmark Recommendation Policies}
\label{sec:benchmarkpolicy}
\noindent
We now characterize benchmark policies. In the scenario of known opportunity costs, we introduce three benchmarks: full transparency, first-best, and second-best, depending on the information design of $\EE_t$. The ARP in Algorithm  \ref{alg:adaptivesampling} gives an example of the second-best policy. Comparing three benchmark policies helps highlight the role of incentives, in addition to the exploration-exploitation tradeoff in classical MAB.

\subsection{Benchmark Policies}
\label{sec:benchmarkpolicies}
\noindent
We first characterize benchmark policies assuming that each agent's opportunity cost is known to the designer. 
\paragraph{\emph{Full Transparency.}} The designer truthfully discloses all her information to the agents \change{(see, e.g., \citet{immorlica2020incentivizing})}. 
Under the assumption in \eqref{eqn:positiveplat}, all her information is
$\EE_t =\{\mu_1>c_*\} \cup \left\{(I_s,y_s,X_{I_s,s}): 1\leq s\leq t-1\right\}$.
Let  $a_t^* = \arg\max_{i\in\AA}\{\mu_1:i=1\}\cup\{\hat{\mu}_{i,t}:i\neq 1\}$, where 
\begin{equation*} 
\hat{\mu}_{i,t}= \frac{\sum_{s\leq t-1:I_s=i}X_{i,s}\cdot y_s}{|\{s\leq t-1:I_s=i, y_s=1\}|}, \quad \forall i\neq 1.
\end{equation*}  
The full transparency can be implemented using the policy in Eq.~\eqref{eqn:randrecpolicy} with $p_{a_i^*,t}=\mathbf{1}\{\hat{\mu}_{a_t^*,t}\geq c_*\}.$
Conditional on the assumption in Eq.~\eqref{eqn:positiveplat},  full transparency would recommend arm $a_t^*=1$ to all agents, as the designer has no prior information on other arms.
This policy fulfills the designer's exploitation goal of minimizing the short-term regret of agents. 

\paragraph{\emph{First-Best Policy.}} The designer optimizes her policy $p_{i,t}$ to minimize $R_T$ in Eq.~\eqref{eqn:defregret}. By ignoring the incentive constraint in Eq.~\eqref{eqn:incentivesct}, the first-best policy captures the classical exploration-exploitation tradeoff, as studied in a rich literature of MAB \citep{rothschild1974two,lai1985asymptotically, even2006action, bubeck2012}. 

Here the first-best policy is implemented in the context of a multi-agent MAB problem. An example of the first-best policy can be implemented by the Active Arms Elimination algorithm proposed by \cite{even2006action}. Note that we provided a version of this algorithm with the incentive guarantee in Section \ref{sec:ARP}. Here each arm is initially labeled active and is included in a set $\BB$. Let $\hat{\mu}_i^l$ be the empirical mean reward of arm $i$ based on $l\geq 0$ rewards, where $\hat{\mu}_i^0=\mu_1$ for any $i\in\AA$.
At each step $l\geq 1$, we set $\hat{\mu}^{l}_* = \max_{i\in\BB}\hat{\mu}_i^{l}$. An arm $i$ is suboptimal if the following condition holds,
\begin{equation*}
\hat{\mu}_i^l+\sqrt{\frac{\ln(T\theta_\tau)}{2l}}<\hat{\mu}_*^l,
\end{equation*} where $\theta_\tau$ is defined as in Eq.~\eqref{eqn:thetatau}.
The suboptimal arms are permanently eliminated from $\BB$.
For the arriving agents, the designer chooses an arm  $i\in\BB$ in lexicographic order and 
recommends it to agents \emph{until} it collects a new reward of arm $i$. 
This step concludes when the designer collects new rewards of all arms in $\BB$, and the algorithm proceeds to step $l+1$. The arms drop out over time until only one arm remains in $\BB$, and the designer commits to exploit the single arm in $\BB$. It is known that this algorithm enjoys minimax optimality in regret if the incentive constraint in Eq.~\eqref{eqn:incentivesct} is ignored \citep{even2006action}.
Comparing first-best and full transparency thus highlights the designer's exploration goal.

\paragraph{\emph{Second-Best Policy.}} In this regime, the designer optimizes her policy $p_{i,t}$ to minimize the regret $R_T$ subject to the incentive constraint in Eq.~\eqref{eqn:incentivesct}. The ARP in Algorithm \ref{alg:adaptivesampling} is a second-best policy. Comparing second-best and first-best policies highlights the role of incentives.

\subsection{The Role of Incentives}
\noindent
We now discuss ARP as an example of the second-best policy and highlight the role of incentives. 
Theorem \ref{thm:knowncost} guarantees the agent's incentive constraint if the exploration rate for arm $i>1$ is no greater than $p_{i,t} $ specified by Eq.~\eqref{eqn:choiceofLi}. That is,
\begin{equation*}
p_{i,t} = \frac{\lambda}{2(c_*-\widehat{M}_i)+\lambda} \ \text{ if } \widehat{M}_i< c_*,
\end{equation*}
and $p_{i,t}=1$ if $\widehat{M}_i\geq c_*$. 
In words, $p_{i,t}$ is the maximum exploration that the designer can recommend, subject to the  incentive constraint. We thus interpret $p_{i,t}$ as the designer's \emph{exploration capacity}.

The capacity depends on the historical rewards via $\widehat{M}_i$. If $\widehat{M}_i\geq c_*$, then the agents have myopic incentives to explore, and the designer can employ the full exploration at $p_{i,t}=1$ for  arm $i$. Therefore,  the incentive constraint in Eq.~\eqref{eqn:incentivesct}  is never binding in this case. 

\begin{figure}[t!]
\centering
\includegraphics[width=0.7\textwidth]{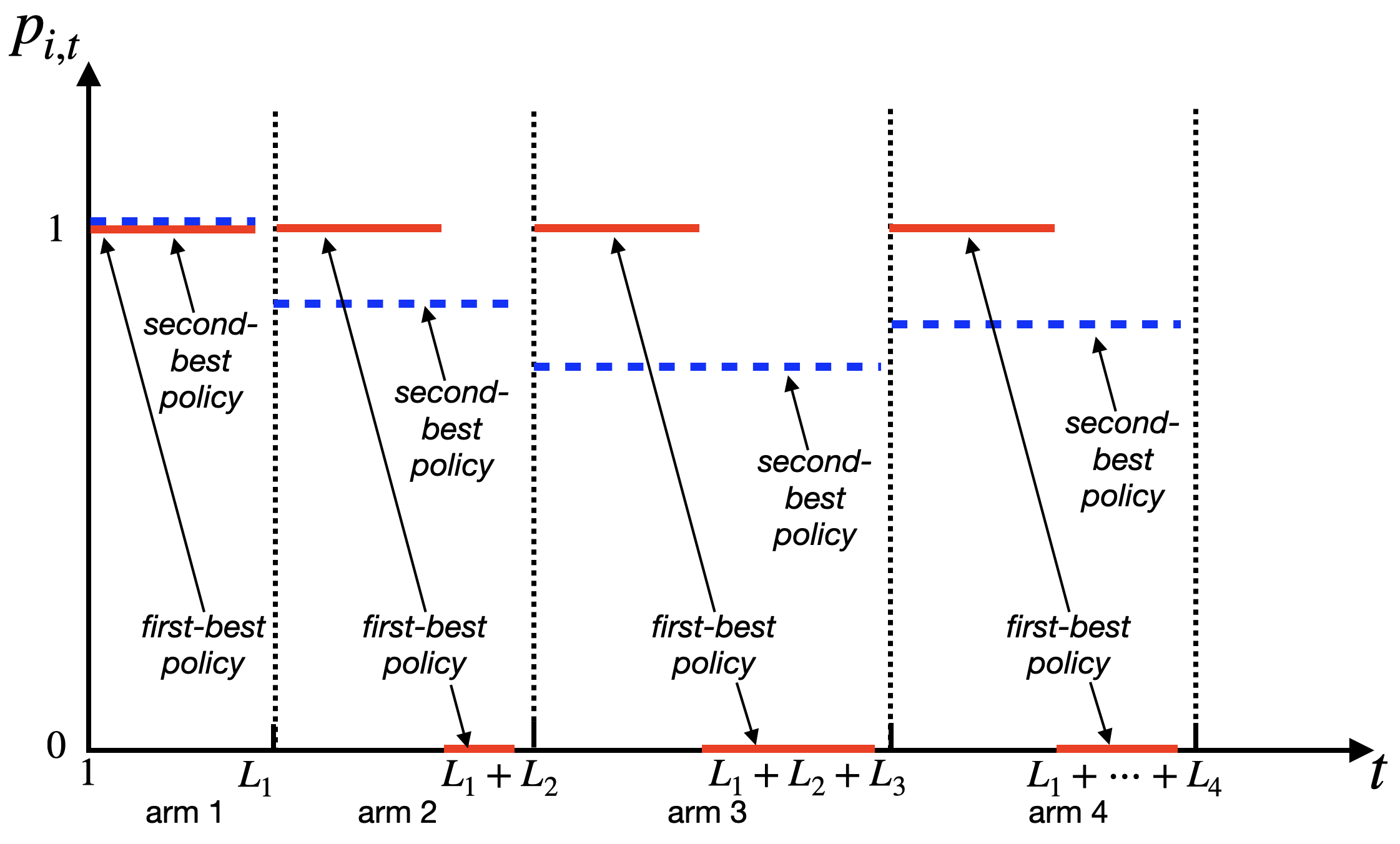}
\caption{Illustrating paths of $p_{i,t}$ for first-best and second-best policies.}
\label{fig:pathofexplore}
\end{figure}

In contrast, if $\widehat{M}_i< c_*$, the incentive constraint  is binding. In this case, the exploration capacity  $p_{i,t}<1$, so not all agents are recommended an explore arm. Intuitively, if the designer recommends an explore arm $i$ to all agents at stage $i$ (i.e., $p_{i,t}=1$), the agents will find the recommendation completely uninformative; therefore, their expected reward of the recommended arm equals $\widehat{M}_i$. Since $\widehat{M}_i<c_*$, they will never pull the explore arm $i$. Moreover, if the designer rarely explores (i.e., $p_{i,t}\approx 0$), then the agent's expected reward of the recommended arm will be at least $\mu_1$;  since $\mu_1>c_*$, the agents will be almost certain that the recommendation is genuine. Naturally, there is an interior level of exploration that will satisfy the incentive constraint. 
The capacity $p_{i,t}$ is initially one for $i=1$. After exploiting arm $i=1$ with $\mu_1>c_*$,
the designer has built her credibility. As time progresses, the exploration capacity $p_{i,t}$ for each arm $i>1$ will be adaptive to historical rewards. 

In essence, the designer pools recommendations across two very different circumstances in ARP: the exploit arm $a_i^*$, and an explore arm $i$, where $i>1$. Although the agents in the latter case will never knowingly follow the recommendation, pooling the two circumstances for recommendations enables the designer to incentivize the agents to explore, conditional on the designed public information that the recommendation in the latter circumstance is kept sufficiently infrequent. Because the agents do not internalize the benefits of exploration, such pooling becomes a useful tool for the designer's second-best policy, as implemented in ARP. 

Figure \ref{fig:pathofexplore} depicts the learning trajectories of the first-best and second-best policies, which have a cutoff structure with a maximal feasible exploration. The maximal exploration equals $1$ under the first-best policy in Section \ref{sec:benchmarkpolicies}, conditional on the corresponding arm remaining active in set $\BB$. Otherwise, no exploration is chosen.  The maximal feasible exploration is $p_{i,t}$ under the second-best policy for each arm $i>1$, where $p_{i,t}$ is adaptive to  historical rewards. Throughout, $p_{i,t}$ remains $1$ or strictly below $1$. In other words, learning is slower under the second-best policy than that under the first-best policy. The designer needs to experiment longer under the second-best regime than under the first-best regime. 

\subsection{Cold-Start Problem in Second-Best Policy}
\noindent
Some products (e.g., songs, movies, or books) are relatively unappealing ex ante, so few people will find them worthwhile to explore on their own, even at a zero price. However, exploring these products can be valuable as some are ultimately worth consumption, and the exploration will benefit subsequent users. The lack of sufficient initial discovery is often coined as the \emph{cold-start} problem, which may lead to the demise of good products \citep{lika2014facing}. The challenge lies in that users who explore the products usually do not internalize the benefit accruing to future users. Agents are myopic and choose actions that satisfy the incentive constraint in Eq.~\eqref{eqn:incentivesct}.

\begin{figure}[t!]
\centering
\includegraphics[width=\textwidth]{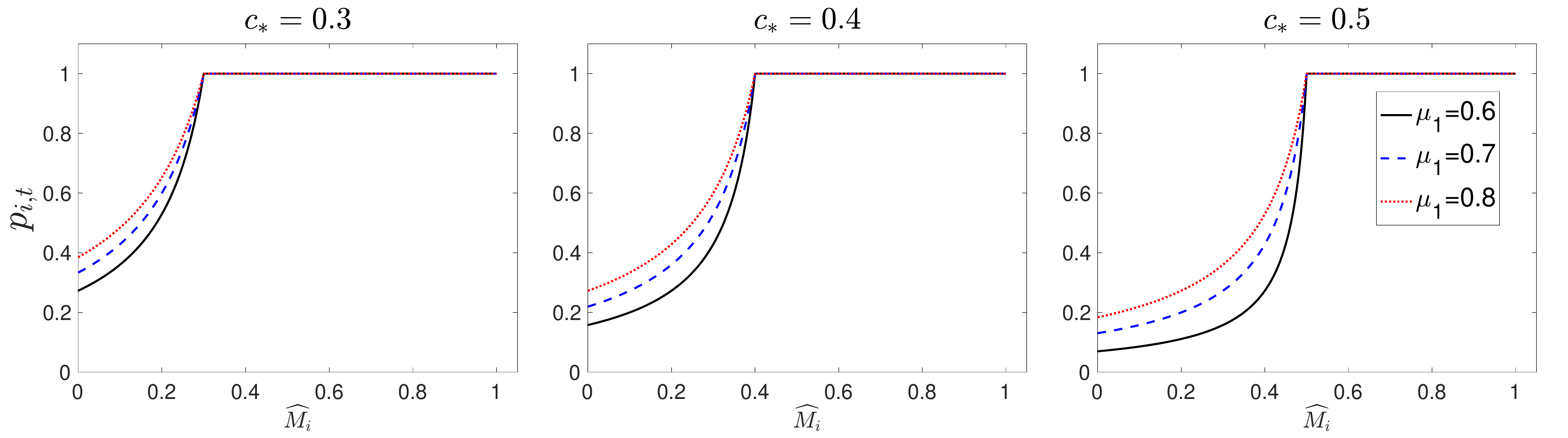}
\caption{(Second-Best) The exploration rate $p_{i,t} $ in Eq.~\eqref{eqn:choiceofLi} for arm $i>1$.}
\label{fig:cold-start}
\end{figure}

The values of $(\mu_1,c_*)$ parameterize the severity of the cold-start problem. 
Figure \ref{fig:cold-start} shows the exploration rate $p_{i,t} $ in Eq.~\eqref{eqn:choiceofLi} for arm $i>1$. 
As $\mu_1$ decreases or $c_*$ increases, it is more difficult for the designer to credibly explore arms, thereby reducing the exploration rate for ARP and making the cold-start problem more severe. 
On the other hand, when the difference $(\mu_1-c^*)$ increases, the sample size $k$ in Eq.~\eqref{eqn:choiceofk} can be smaller for the initial collection of each arm's rewards. Moreover, the margin parameter  $\lambda$ in Eq.~\eqref{eqn:choiceoflambda} can be larger, which leads to an increased exploration rate for ARP and makes the cold-start problem less severe. 

Background learning for $\mu_1$ seeds the exploration. In particular, the assumption in Eq.~\eqref{eqn:positiveplat} would fail if no arm's expected reward is known or $\mu_1<c_*$. In either of these cases, the designer has no credibility to induce exploration. This observation has practical implications. For instance, Internet platforms such as Pandora make costly investments to discover $\mu_1$, which helps speed up learning and discovery of good products in the second-best regime.

\section{Optimal Policy Under Private Opportunity Costs}
\label{sec:privatereveal}

\noindent
We now consider that the designer starts with little information about agents' opportunity costs and directly elicits necessary knowledge of agents' preferences. Each agent's opportunity cost can be \emph{private}, heterogenous, and unknown to the designer.  
In this case, the incentive constraint in Eq.~\eqref{eqn:incentivesct} cannot always be satisfied due to the unknown opportunity costs. \change{However, since agents share the same utility and their actions follow the incentive constraint in Eq.~\eqref{eqn:incentivesct}, their actions can be used as feedback  to improve the exploration process and to identify the optimal arm for the agents.} This section discusses a second-best policy that satisfies ex-post fairness and achieves minimax regret $R_T$ under private opportunity costs.

\subsection{Modified Adaptive Recommendation Policy}
\label{sec:knownrounds}
\noindent 
We propose a \emph{modified adaptive recommendation policy} (MARP), which is also built upon ``randomized recommendation" and ``adaptive exploration.”  MARP has two new ingredients compared to ARP in Algorithm \ref{alg:adaptivesampling}. First, ARP considers only two arms  for the recommendation at each round, an explore arm $i$ and an exploit arm $a_i^*$, and  $p_{j,t} = 0$ for any $j\in\AA\setminus\{ a_i^*,i \}$, as specified in Eq.~\eqref{eqn:choiceofLi}. In contrast, MARP considers all arms for the recommendation at each round, where $p_{j,t} \neq 0$ for any $j\in\AA$. Second, MARP uses each arm's cumulative loss as feedback to adjust the probability of recommending each arm, whereas ARP uses the empirical mean of historical rewards as a factor to adjust the probability of recommending each arm. 

\begin{algorithm}
\caption{ \normalsize{{Modified Adaptive Recommendation Policy (MARP)}}}
\label{alg:MARP}
\begin{algorithmic}[1]
\State  \normalsize{\textbf{Input:} Arm set $\AA$; Parameter: $\eta=\sqrt{8(\ln m)/T}$.}
\State \textbf{Initialization:}  Initialize the cumulative loss $\hat{L}_{i,0}=0,\forall i\in\AA$. Announce $\EE_1$ according to Eq.~\eqref{eqn:publiceetinitialagent}. Choose an arm $i\in\AA$ with $p_{i,1}=1/m$ and recommend it to agent $t=1$. Collect a reward if $y_1=1$. 
\State \textbf{for} each agent $t>1$  in increasing lexicographic order   \textbf{do}
\State \quad\quad For each arm $i\in\AA$, compute the estimated loss $\hat{l}(i,y_{t-1})$ in Eq.~\eqref{eqn:estimatedloss};
\State \quad\quad Update the estimated cumulative loss $\hat{L}_{i,t-1} = \hat{L}_{i,t-2}+\hat{l}(i,y_{t-1})$;
\State \quad\quad Update the vector $\bp_{t}$ by
$p_{i,t} = \frac{\exp(-\eta\cdot\hat{L}_{i,t-1})}{\sum_{i=1}^m\exp(-\eta\cdot\hat{L}_{i,t-1})},\forall i\in\AA$ in Eq.~\eqref{eqn:expweightstrategypractice};
\State \quad\quad Update $\EE_t$ according to Eq.~\eqref{eqn:updatedeetunknowncost};
\State \quad\quad Recommend $I_t$ in Eq.~\eqref{eqn:randrecpolicyindfairt} to agent $t$. Collect a reward for arm $I_t$ if $y_t=1$.
\State \textbf{end}
\end{algorithmic}
\end{algorithm}

We summarize MARP in Algorithm \ref{alg:MARP} and then detail its main steps.
The first step is the input of parameter $\eta>0$. Consider that the time horizon $T$ is known. Let 
\begin{equation*}
\eta=\sqrt{\frac{8\ln m}{T}},
\end{equation*}
where $m$ is the number of arms. We will extend the algorithm to the case of unknown $T$ later in this section.

The second step is the \emph{exploration-exploitation} of arms. 
Since agents' opportunity costs are heterogeneous and unknown, their actions vis-a-vis recommendations cannot directly yield an inference of the optimal arm up to a given round. For example, an arm $i\in\AA$ whose recommendation is ignored by an agent $j$ does not necessarily have a worse expected reward than another arm $i'\in\AA$ whose recommendation is followed by another agent $j'$, where $i\neq i'$ and $j\neq j'$. We use a strategy of randomly pooling recommendations across all arms, rather than pooling from only a genuinely good arm and an unknown arm as used in ARP. This approach prevents the optimal arm from being eliminated early. Specifically, for agent $t=1$, the designer sends the following information,
\begin{equation}
\label{eqn:publiceetinitialagent}
\begin{aligned}
\EE_1 & =  \left\{p_{i,1} = \frac{1}{m}, \ \forall i\in\AA\right\}.
\end{aligned}
\end{equation}
Then the designer recommends arm $I_1$ to agent $t=1$ according to the policy in Eq.~\eqref{eqn:randrecpolicy}, where the vector $\bp_t$   is given by $p_{i,1}=1/m$ for any $i\in\AA$. 
Conditional on $\EE_1$ in Eq.~\eqref{eqn:publiceetinitialagent}, agent $t=1$ is aware of the randomized policy. 
The designer collects a reward $X_{I_1,1}$ if  $y_1=1$, that is, if agent $t=1$ pulls the recommended arm $I_1$. 

Next, we consider agent $t>1$. Here the designer considers all arms for the recommendation, where the $\bp_t$ in Eq.~\eqref{eqn:randrecpolicy}  is specified as,
\begin{equation}
\label{eqn:expweightstrategy}
p_{i,t} = \frac{\exp\left(-\eta\cdot\sum_{s=1}^{t-1}l(i,y_s)\right)}{\sum_{j=1}^m\exp\left(-\eta\cdot\sum_{s=1}^{t-1}l(j,y_s)\right)},\quad \text{for }i\in\AA.
\end{equation}
The $\eta$ is a parameter with $\eta>0$, and $l(i,y_s)$ is the loss function defined in Eq.~\eqref{eqn:loss}. 
It is seen that MARP uses the agent's action $y_s$ and loss $l(i,y_s)$ as feedback to update $p_{i,t}$. 
Moreover, $p_{i,t}$ in Eq.~\eqref{eqn:expweightstrategy} is adaptive to the historical data through the cumulative loss $\sum_{s=1}^{t-1}l(i,y_s)$ for each $i\in\AA$. This is intuitive because $p_{i,t}$ should be large for an arm $i$ with a small cumulative loss. In contrast,  ARP in Algorithm \ref{alg:adaptivesampling} is adaptive to the historical rewards through the empirical mean of the historical rewards $\widehat{M}_i$ in Eq.~\eqref{eqn:choiceofLi}. Note that $\widehat{M}_i$ does not differentiate each arm's individual rewards. Thus, the adaptivity of MARP is different from that of ARP.
An explanation of this difference is that if the opportunity costs are known, ARP can infer an optimal arm conditional on the information up to a given round and then uses this arm as an exploit arm. 
From the agent's perspective, they can infer the quality of this exploit arm through the empirical mean of historical rewards, which directly determines the largest exploration rate in ARP. 
However, given private opportunity costs, an agent's action $y_s$ towards an arm $i\in\AA$ provides little insight into the other agents' actions $y_{s'}$ towards the same arm $i\in\AA$.  As a solution, MARP traces down the individual loss of each arm $i\in\AA$ and gradually improves recommendations by adjusting $p_{i,t}$ over time.

We consider a practical approach to compute $p_{i,t}$ in Eq.~\eqref{eqn:expweightstrategy}. Note that $p_{i,t}$ remains unchanged if we let $l(i,y_s) = \E[-X_{I_s,s}\cdot y_s]$, where the expectation is taken over randomized $I_t$ and stochastic reward $X_{I_t,t}$.
An estimator for $l(i,y_s)$ is given by, 
\begin{equation}
\label{eqn:estimatedloss}
\hat{l}(i,y_s)=
\begin{cases}
-X_{i,s}\cdot y_s/p_{i,s}, & \text{ if } I_s=i,\\
0, & \text{ if } I_s \neq i,
\end{cases}\quad \forall i\in\AA \text{ and } 1\leq s\leq t-1,
\end{equation}
where $X_{i,s}$ is the collected reward if $y_s=1$. Then $\hat{l}(i,y_s)$ is unbiased since $\E\left[\hat{l}(i,y_s)\right] = \sum_{j=1}^mp_{j,s}\E\left[-X_{i,s}\cdot y_s/p_{i,s}\right]\mathbf{1}_{j=i} = \E\left[-X_{I_s,s}\cdot y_s\right]$.
Hence we can update $p_{i,t}$ in Eq.~\eqref{eqn:expweightstrategy} using the estimator $\hat{l}(i,y_s)$,
\begin{equation}
\label{eqn:expweightstrategypractice}
p_{i,t} = \frac{\exp\left(-\eta\cdot\sum_{s=1}^{t-1}\hat{l}(i,y_s)\right)}{\sum_{j=1}^m\exp\left(-\eta\cdot\sum_{s=1}^{t-1}\hat{l}(j,y_s)\right)},\quad \text{for }i\in\AA.
\end{equation}
Next, the message $\EE_t$ for agent $t>1$  is updated as,
\begin{equation}
\label{eqn:updatedeetunknowncost}
\begin{aligned}
\EE_t  = \left\{p_{i,t}\propto \text{the exponential of arm $i$'s cumulative reward,} \ \forall i\in\AA \right\}\cup\left\{X_{I_s,s}: y_s=1 \text{ and } 1\leq s\leq t-1\right\}.
\end{aligned}
\end{equation}
Under $\EE_t$ in Eq.~\eqref{eqn:updatedeetunknowncost}, agents do not observe past agents' actions or the identities of the arms that were recommended to past agents. Instead, $\EE_t$ is transparent in disclosing the history of rewards received by past agents who have followed the recommendations.

The designer can incorporate each agent's belief  $\HH(t)$ into the recommendation. In particular, the designer can employ  the following policy,
\begin{equation}
\label{eqn:randrecpolicyindfairt}
 I_t = ~  \sum_{j\in\AA} j\cdot \mathbf{1}\left\{U_t\in\left[\sum_{s=1}^{j-1}p_{s,t},\sum_{s=1}^jp_{s,t}\right)\right\}\cdot \mathbf{1}\{\HH(t) \text{ holds}\}  +  \left[\underset{j\in\AA}{\arg\min}\sum_{s=1}^{t-1}\hat{l}(j,y_s)\right]\cdot\mathbf{1}\{\HH(t) \text{ fails}\}.
\end{equation}
Here $\hat{l}(i,y_s)$ and $p_{s,t}$ are given by Eq.~\eqref{eqn:estimatedloss} and Eq.~\eqref{eqn:expweightstrategypractice}, respectively.  This policy is a generalization of the randomized recommendation in Eq.~\eqref{eqn:randrecpolicy}, with an additional  criterion $\HH(t)$.
After recommending an arm to agent $t$, the designer collects a reward $X_{I_t,t}$ if $y_t=1$.  

\subsection{Performance Analysis for MARP} 
\label{sec:indfairnessmarp}
\noindent
We provide the performance analysis for MARP in Algorithm \ref{alg:MARP}. First, 
we show that although now agents' opportunity costs are unknown, MARP satisfies ex-post fairness defined in Eq.~\eqref{def:expostfair}.

\begin{theorem}
\label{lem:individualfairmarp2}
The MARP  procedure in Algorithm \ref{alg:MARP} satisfies the ex-post fairness criterion in Eq.~\eqref{def:expostfair}.
\end{theorem}

Next, we provide the regret analysis for the MARP procedure. The following theorem establishes a nonasymptotic regret bound.
\begin{theorem}
\label{thm:regretexpweight}
If the parameters $\eta=\sqrt{(8\ln m)/T}$ and $\gamma_t\equiv 1$ in Eq.~\eqref{eqn:egofht}, then for any arm set $\AA$ and action sequence $\{y_t\}_{t=1}^T\in\YY^T$, the MARP procedure in Algorithm \ref{alg:MARP} satisfies,
with probability at least $1-\delta$ for any $\delta\in(0,1)$,
\begin{equation*}
R_T\leq \sqrt{\frac{T}{2}\ln m} +  \sqrt{\frac{T}{2}\ln\frac{1}{\delta}},
\end{equation*}
where $R_T$ is defined in Eq.~\eqref{eqn:defregret}, $T$ is the time horizon, and $m$ is the number of arms.
\end{theorem}
\noindent
To minimize the regret in Eq.~\eqref{eqn:defregret}, it is clear that any deterministic policy of the designer is insufficient. This is because there exists an action sequence  $\{y_t\}_{t=1}^T$ such that the designer has the loss, $l(I_t,y_t)=1$, at every time instant $t$.  However, the randomization in Eq.~\eqref{eqn:randrecpolicyindfairt} guarantees the sublinear regret bound in terms of $T$ even when the loss function $l(\cdot,\cdot)$ in Eq.~\eqref{eqn:loss} is nonconvex, as shown in Theorem \ref{thm:regretexpweight}.
Moreover, using the classical notion of \emph{Hannan consistency} in games \citep{hannan1957approximation},
it is clear that  MARP  is Hannan consistent as $\lim\sup_{T\to\infty} R_T/T=0$.

Next, we establish a corresponding lower bound for regret. Define the minimax regret for this problem as
\begin{equation*}
R^\dagger_T\equiv  \inf_{\{I_t\}_{t=1}^T}\sup_{\{\AA:|\AA|=m\}, \{y_t\}_{t=1}^T\in\YY^T} R_T,
\end{equation*}
where the infimum is taken over all possible recommendation policies $\{I_t\}_{t=1}^T$, and the supremum is taken over all possible classes of $m$ arms and sequences of $T$ actions.  
Then $R^\dagger_T$ measures the best possible performance of a recommendation policy.  
\begin{theorem}
\label{thm:lowerbd}
For any arm set $\AA$ and  action sequence $\{y_t\}_{t=1}^T\in\YY^T$, the minimax regret $R^\dagger_T$ satisfies
\begin{equation*}
\lim_{m\to\infty}\lim_{T\to\infty}\frac{R^\dagger_T }{(1-\epsilon_T)\sqrt{T\ln m/2}}=1,
\end{equation*}
where $\epsilon_T>0$  is  any parameter with $\lim_{T\to\infty}\epsilon_{T}=0$.
\end{theorem}
\noindent
This theorem implies that the upper regret bound by MARP  in Theorem \ref{thm:regretexpweight} is essentially unimprovable. 
On the other hand, ARP also achieves a comparable bound of $O(\sqrt{T})$ in Theorem \ref{lem:regretofknowncost}. 
Although the two algorithms achieve similar levels of regret, there exists a key difference between them due to the tradeoff of knowing additional information about opportunity costs and the incentive guarantee. First, ARP requires knowledge of the opportunity costs, which is unnecessary for implementing MARP. Second, ARP guarantees the agent's incentive as shown in Theorem \ref{thm:knowncost}, while MARP does not enjoy such a guarantee. 

\subsection{Unknown Round Information}
\label{sec:unknownrounds}

\noindent 
The MARP algorithm has the disadvantage that it requires knowledge of the time horizon $T$ in advance. 
Hence the result in Theorem \ref{thm:regretexpweight} does not hold uniformly over sequences of agents with any length but only for sequences of agents with a given length $T$. However, we show that if MARP in Algorithm \ref{alg:MARP} is equipped with a time-varying parameter $\eta=\sqrt{8(\ln m)/t}$, then it yields a near-optimal regret bound for any unknown $T$. The following theorem applies to the nonconvex loss $l$ and the estimated loss $\hat{l}$ in Algorithm \ref{alg:MARP}. 

\begin{theorem}
\label{thm:nearoptimaldivergingT}
If the parameter $\eta$ is chosen as $\eta_t=\sqrt{(8\ln m)/t}$ for $t\geq 1$, then for any arm set $\AA$ and action  sequence $\{y_t\}_{t=1}^T\in\YY^T$, the MARP procedure in Algorithm \ref{alg:MARP} satisfies, with probability at least $1-\delta$ for any $\delta\in(0,1)$,
\begin{equation*}
R_T\leq \sqrt{2T\ln m} +\sqrt{\frac{\ln m}{8}} +  \sqrt{\frac{T}{2}\ln\frac{1}{\delta}},
\end{equation*}
where $R_T$ is defined in Eq.~\eqref{eqn:defregret}, $T$ is the time horizon, and $m$ is the number of arms.
\end{theorem}

\section{Numerical Examples}
\label{sec:numericalexamples}

\noindent
\change{In this section, we provide a numerical investigation of the empirical performance of ARP and MARP in Algorithms \ref{alg:adaptivesampling} and \ref{alg:MARP}, respectively. We compare ARP and MARP with alternative methods, including the upper confidence bound algorithm (UCB) \citep{lai1985asymptotically}, the elimination algorithm \citep{even2006action}, and Thompson sampling \citep{thompson1933likelihood}. As described in Algorithm \ref{alg:adaptivesampling}, ARP explores each arm while conducting sufficient exploitations to keep agents incentivized during sampling. After collecting enough information on each arm, it eliminates suboptimal arms by gradually removing those with lower empirical rewards. The incentive is maintained since agents do not prefer arms with low rewards. 
As described in Algorithm \ref{alg:MARP}, MARP performs adaptive recommendations based on the individual loss of each arm, recommending low-loss arms and reducing exposure to high-loss arms. 
The incentive is incorporated since agents without incentive create the highest loss, reducing the exposure of arms that fall below the agents' incentive.
The Elimination algorithm rules out arms with lower empirical rewards regardless of agents' incentives and without collecting any prior arm information.
The UCB calculates confidence bounds for each arm using historical rewards and selects the arm with the highest upper confidence bound, even though the corresponding rewards may be lower than the opportunity costs. In such cases, agents may not follow the recommendation.
Thompson sampling adjusts recommendation probabilities iteratively, trying to increase the exposure of arms already pulled by incentivized agents.}

\subsection{Example on Gaussian Rewards}
\label{sec:gaussianbandit}
\noindent
\change{The first example studies the case where the stochastic reward of the $t$-th agent pulling arm $i$, $X_{i,t}$, is drawn i.i.d. from a Gaussian distribution. Specifically, $X_{i,t}$ follows $\mathcal{N}(\mu_i, 0.1^2) \cap [0, 1]$, where $\{\mu_i\}_{i=1}^{m}$  is drawn uniformly from $[0, 0.6]$. 
Suppose there are varying numbers of arms, $m = \{5, 10, 15\}$. We consider two settings: for ARP, the opportunity costs are public and uniform, $c_t = c_*$, $c_* \in \{0.20, 0.25, 0.30\}$; for MARP, $c_t$ follows a Beta distribution in $\{Beta(1, 2), Beta(1, 2.5), Beta(1, 3)\}$, unknown to the designer.
To implement ARP, we set $k = 10$, and choose $\lambda$ from $[0, 1]$ without further restrictions,  while $\tau$ only needs to lie in $(0, 1 - c_*)$. We set $\tau = 0.2$ and compute $\theta_\tau$ in Eq.~\eqref{eqn:thetatau}, noting that $\min_{i \in \mathcal{A}} \mathbb{P}(\mu_i - c_ \geq \tau) = 1 - 5/3 \cdot (c_* + \tau)$. To ensure constraint Eq.~\eqref{eqn:publiceetinitial},  we additionally set $\mu_1$ to $c_*$ in ARP. To implement the elimination algorithm, we use Algorithm 3 in \citep{even2006action} with $c = 10$ and $\delta = 0.05$. Parameters in the other methods can be calculated accordingly.
The data are simulated by having $T = 5000$ agents receive recommendations on arms.  We summarize results based on $500$ data realizations.

\begin{table}[thb!]
\centering
\caption{The mean regret and the $90\%$ confidence interval of ARP and alternative algorithms for Section \ref{sec:gaussianbandit}, based on 500 data replications. Boldface indicates the method with the smallest mean regret.}
\resizebox{\textwidth}{!}{
\begin{tabular}{cccccccccccc}
\toprule
& \multicolumn{3}{c}{$m=5$} && \multicolumn{3}{c}{$m=10$} && \multicolumn{3}{c}{$m=15$} \\ \cline{2-4} \cline{6-8} \cline{10-12}
& $c_* = 0.20$ & $c_* = 0.25$ & $c_* = 0.30$ && $c_* = 0.20$ & $c_* = 0.25$ & $c_* = 0.30$ && $c_* = 0.20$ & $c_* = 0.25$ & $c_* = 0.30$ \\ [0.2ex]
 \midrule
ARP & $\textbf{159.51}$ & $\textbf{165.87}$ &  $\textbf{172.94}$ && $\textbf{323.52}$ & $\textbf{330.89}$ & $\textbf{341.50}$ && $\textbf{466.19}$ & $\textbf{475.08}$ & $\textbf{484.85}$ \\
 & $\textbf{(83.88, 253.58)}$ & $\textbf{(90.60, 255.41)}$ & $\textbf{(97.70, 258.63)}$ && $\textbf{(233.02, 416.55)}$ & $\textbf{(235.32, 423.53)}$ & $\textbf{(251.74, 435.40)}$ && $\textbf{(361.71, 560.32)}$ & $\textbf{(371.77, 569.57)}$ & $\textbf{(384.44, 579.24)}$ \\ [0.2ex]
Elimination &  $634.21$ & $982.83$ &   $1401.45$ && $547.65$ & $892.02$ &   $1594.64$ && $526.27$ & $956.01$ &  $1610.82$ \\
 & $(193.72, 2090.84)$ & $(215.88, 2394.00)$ & $(246.96, 2564.48)$ && $(305.80, 1817.06)$ & $(318.40, 2535.76)$ & $(346.09, 2661.75)$ && $(344.03, 626.93)$ & $(354.83, 2593.17)$ & $(369.99, 2661.40)$ \\ [0.2ex]
UCB & $462.19$ & $723.31$ &  $1233.74$ && $424.12$ & $767.30$ & $1410.50$ && $486.78$ & $657.80$ & $1468.38$ \\
 & $(119.87, 2034.09)$ & $(122.61, 2307.06)$ & $(132.54, 2514.20)$ && $(246.88, 1769.08)$ & $(248.70, 2578.51)$ & $(256.53, 2648.77)$ && $(322.63, 503.47)$ & $(328.75, 2488.87)$ & $(333.02, 2666.94)$ \\ [0.2ex]
Thompson & $1206.96$ & $1438.11$ &   $1653.31$ && $1272.91$ & $1625.28$ & $1943.88$ && $1257.57$ & $1667.16$ & $2065.41$ \\
 & $(395.72, 2418.85)$ & $(400.48, 2522.50)$ & $(388.91, 2605.96)$ && $(605.10, 2461.02)$ & $(631.71, 2644.48)$ & $(579.93, 2671.03)$ && $(762.84, 2335.42)$ & $(775.64, 2649.96)$ & $(749.10, 2675.48)$ \\ [0.2ex]
\bottomrule
\end{tabular}
}
\label{table:Gauss-ARP}
\end{table}
Table \ref{table:Gauss-ARP} reports the mean regret and its $90\%$ confidence interval under different values of $m$ and $c_*$ for ARP, elimination, UCB, and Thompson sampling. We find that ARP consistently exhibits lower mean regret compared to the other algorithms because ARP keeps agents incentivized throughout the whole process, while the Elimination and UCB algorithms explore at any cost, and Thompson sampling is too slow to find the optimal arm. Additionally, ARP has the shortest confidence interval bandwidth. As the number of arms increases, the regret also increases since more rounds are needed to eliminate arms.

\begin{table}[thb!]
\centering
\caption{The mean regret and the $90\%$ confidence interval of MARP and alternative algorithms for Section \ref{sec:gaussianbandit}, based on 500 data replications. Boldface indicates the method with the smallest mean regret.}
\resizebox{\textwidth}{!}{
\begin{tabular}{ccccccccccccc}
\toprule
& \multicolumn{3}{c}{$m=5$} && \multicolumn{3}{c}{$m=10$} && \multicolumn{3}{c}{$m=15$} \\ \cline{2-4} \cline{6-8} \cline{10-12}
& $Beta(1, 2)$ & $Beta(1, 2.5)$ & $Beta(1, 3)$ && $Beta(1, 2)$ & $Beta(1, 2.5)$ & $Beta(1, 3)$ && $Beta(1, 2)$ & $Beta(1, 2.5)$ & $Beta(1, 3)$ \\ [0.2ex]
 \midrule
MARP & $\textbf{738.46}$ & $\textbf{576.97}$ &  $\textbf{452.26}$ && $\textbf{838.58}$ & $\textbf{689.38}$ & $\textbf{617.87}$ && $\textbf{1013.10}$ & $\textbf{815.45}$ & $\textbf{696.22}$ \\
 & $\textbf{(557.04, 941.83)}$ & $\textbf{(408.72, 783.91)}$ & $\textbf{(287.48, 625.97)}$ && $\textbf{(575.35, 1576.52)}$ & $\textbf{(401.69, 1398.15)}$ & $\textbf{(302.30, 1645.50)}$ && $\textbf{(587.33, 2162.32)}$ & $\textbf{(423.25, 1901.01)}$ & $\textbf{(314.73, 1807.43)}$ \\ [0.2ex]
Elimination &  $1054.97$ & $895.32$ &   $781.51$ && $1121.88$ & $984.68$ &   $868.23$ && $1172.17$ & $1019.33$ &  $901.65$ \\
 & $(560.71, 2006.36)$ & $(483.57, 1816.43)$ & $(373.79, 1635.94)$ && $(903.73, 1321.13)$ & $(772.71, 1186.69)$ & $(671.27, 1064.28)$ && $(980.42, 1365.38)$ & $(827.04, 1198.31)$ & $(705.82, 1088.00)$ \\ [0.2ex]
UCB & $832.33$ & $660.98$ &  $548.57$ && $1043.24$ & $859.13$ & $726.49$ && $1184.21$ & $999.49$ & $859.17$ \\
 & $(691.41, 972.94)$ & $(526.47, 829.78)$ & $(417.30, 731.82)$ && $(861.37, 1194.95)$ & $(687.87, 1019.49)$ & $(564.88, 882.30)$ && $(985.53, 1379.77)$ & $(805.15, 1178.42)$ & $(689.84, 1023.98)$ \\ [0.2ex]
Thompson & $1331.78$ & $1236.74$ &  $1106.24$ && $1641.67$ & $1528.70$ & $1412.68$ && $1782.51$ & $1651.43$ & $1549.30$ \\
 & $(725.56, 1981.83)$ & $(653.11, 1981.67)$ & $(520.12, 1796.22)$ && $(1143.63, 2101.52)$ & $(913.44, 2089.61)$ & $(828.25, 1967.55)$ && $(1382.55, 2180.94)$ & $(1231.33, 2086.62)$ & $(1036.43, 2024.78)$ \\ [0.2ex]
\bottomrule
\end{tabular}
}
\label{table:Gauss-MARP}
\end{table}
Table \ref{table:Gauss-MARP} reports the mean regret and its $90\%$ confidence interval under different $m$ values and $c_t$ distributions for MARP, elimination, UCB, and Thompson sampling. MARP consistently exhibits the smallest mean regret among the algorithms.
Moreover, $c_t$ follows a Beta distribution $Beta(\alpha, \beta)$, and an increase in the parameter $\beta$ corresponds to a greater proportion of  agents with lower opportunity costs entering the market. We observe that as  $\beta$ increases, there is a corresponding decrease in the mean regret.  This is due to the fact that lower opportunity costs make it easier to provide incentivized recommendations, which leads to a decrease in regret.}

\subsection{Example on Beta Rewards}
\label{sec:betabandit}
\noindent
\change{The second example considers the case where the stochastic reward $X_{i,t}$ is drawn i.i.d. from $Beta(1, \beta_i)$, and $\{\beta_i\}_{i = 1}^m$ is a random permutation of $\{1, 2, \cdots, m\}$. Then, the expected reward is given by $\mu_i=1/(1+\beta_i)$ for $i=1,2,\ldots,m$. Suppose there are varying numbers of arms $m=\{5, 10, 15\}$. We consider two settings: for ARP, the opportunity costs are public and uniform, $c_t = c_*$, $c_* \in \{0.05, 0.15, 0.25\}$; for MARP, $c_t$ follows a Beta distribution in $\{Beta(1, 2),$ $ Beta(1, 2.5),$ $ Beta(1, 3)\}$ unknown to the designer. To implement ARP, we set $k=10$, $\tau=0.2$ and compute $\theta_\tau$ in Eq.~\eqref{eqn:thetatau}, noting that $\min_{i\in\AA}\P(\mu_i-c_*\geq \tau) = 3/m$ when $c_* = 0.05$, and $\min_{i\in\AA}\P(\mu_i-c_*\geq \tau) = 1/m$ when $c_* = 0.15, 0.25$. To ensure the constraint Eq.~\eqref{eqn:publiceetinitial}, we set $\mu_1$ to $c_*$ in ARP. To implement the elimination algorithm, we use Algorithm 3 in \citep{even2006action} with $c=10$, $\delta=0.05$. Parameters in the other methods can be calculated accordingly. The data are simulated by having $T = 5000$ agents receive recommendations on arms.  We summarize results based on $500$ data realizations.

\begin{table}[thb!]
\centering
\caption{The mean regret and the $90\%$ confidence interval of ARP and alternative algorithms for Section \ref{sec:betabandit}, based on 500 data replications. Boldface indicates the method with the smallest mean regret.}
\resizebox{\textwidth}{!}{
\begin{tabular}{cccccccccccc}
\toprule
& \multicolumn{3}{c}{$m=5$} && \multicolumn{3}{c}{$m=10$} && \multicolumn{3}{c}{$m=15$} \\ \cline{2-4} \cline{6-8} \cline{10-12}
& $c_* = 0.05$ & $c_* = 0.15$ & $c_* = 0.25$ && $c_* = 0.05$ & $c_* = 0.15$ & $c_* = 0.25$ && $c_* = 0.05$ & $c_* = 0.15$ & $c_* = 0.25$ \\ [0.2ex]
 \midrule
ARP & $\textbf{133.90}$ & $\textbf{140.76}$ &  $\textbf{159.53}$ && $\textbf{257.72}$ & $\textbf{273.24}$ & $\textbf{157.21}$ && $\textbf{364.44}$ & $\textbf{386.15}$ & $\textbf{429.42}$ \\
 & $\textbf{(75.07, 247.512)}$ & $\textbf{(81.09, 259.98)}$ & $\textbf{(88.88, 308.92)}$ && $\textbf{(166.38, 455.09)}$ & $\textbf{(182.19, 467.99)}$ & $\textbf{(101.80, 298.87)}$ && $\textbf{(260.23, 544.30)}$ & $\textbf{(263.74, 561.20)}$ & $\textbf{(326.34, 598.19)}$ \\ [0.2ex]
Elimination &  $394.96$ & $578.12$ &   $1427.50$ && $464.62$ & $880.25$ &   $1325.32$ && $494.51$ & $1683.95$ &  $2199.47$ \\
 & $(238.01, 467.23)$ & $(248.11, 2248.60)$ & $(309.90, 2248.63)$ && $(332.63, 572.48)$ & $(352.91, 2248.00)$ & $(112.80, 4452.61)$ && $(371.76, 606.44)$ & $(437.01, 2247.72)$ & $(1497.87, 2247.70)$ \\ [0.2ex]
UCB & $173.65$ & $430.30$ &  $1338.07$ && $321.89$ & $850.78$ & $1210.42$ && $458.01$ & $1427.60$ & $2199.50$ \\
 & $(140.63, 199.61)$ & $(147.72, 2248.60)$ & $(158.73, 2248.63)$ && $(296.48, 375.41)$ & $(302.07, 2248.03)$ & $(166.21, 4452.61)$ && $(433.97, 501.20)$ & $(445.25, 2247.72)$ & $(1497.87, 2247.70)$ \\ [0.2ex]
Thompson & $999.08$ & $1122.10$ &   $1706.69$ && $1310.68$ & $1620.13$ & $3785.23$ && $1527.75$ & $2061.75$ & $2199.52$ \\
 & $(441.59, 1515.62)$ & $(442.13, 2248.60)$ & $(610.43, 2248.63)$ && $(750.99, 1614.28)$ & $(974.57, 2248.08)$ & $(2236.63, 4452.61)$ && $(1020.26, 1764.20)$ & $(1383.62, 2247.72)$ & $(1497.87, 2247.70)$ \\ [0.2ex]
\bottomrule
\end{tabular}
}
\label{table:Beta-ARP}
\end{table}

Table \ref{table:Beta-ARP} reports the mean regret and its $90\%$ confidence interval under different values of $m$ and $c_*$ for ARP, elimination, UCB and Thompson sampling.  We find that ARP consistently outperforms other algorithms in terms of regret across different settings. As $c_*$ increases, the confidence intervals for alternative methods widen significantly, whereas ARP maintains a shorter bandwidth, which shows its robustness in scenarios with higher opportunity costs $c_*$.

\begin{table}[thb!]
\centering
\caption{The mean regret and the $90\%$ confidence interval of MARP and alternative algorithms for Section \ref{sec:betabandit}, based on 500 data replications. Boldface indicates the method with the smallest mean regret.}
\resizebox{\textwidth}{!}{
\begin{tabular}{ccccccccccccc}
\toprule
& \multicolumn{3}{c}{$m=5$} && \multicolumn{3}{c}{$m=10$} && \multicolumn{3}{c}{$m=15$} \\ \cline{2-4} \cline{6-8} \cline{10-12}
& $Beta(1, 2)$ & $Beta(1, 2.5)$ & $Beta(1, 3)$ && $Beta(1, 2)$ & $Beta(1, 2.5)$ & $Beta(1, 3)$ && $Beta(1, 2)$ & $Beta(1, 2.5)$ & $Beta(1, 3)$ \\ [0.2ex]
 \midrule
MARP & $\textbf{799.60}$ & $\textbf{635.72}$ &  $\textbf{513.67}$ && $\textbf{965.53}$ & $\textbf{778.45}$ & $\textbf{663.50}$ && $\textbf{1028.52}$ & $\textbf{885.58}$ & $\textbf{718.33}$ \\
 & $\textbf{(637.49, 1494.76)}$ & $\textbf{(494.96, 1173.76)}$ & $\textbf{(363.64, 1118.18)}$ && $\textbf{(685.54, 1956.31)}$ & $\textbf{(531.88, 1761.01)}$ & $\textbf{(401.46, 1803.52)}$ && $\textbf{(714.62, 1999.52)}$ & $\textbf{(563.51, 2010.01)}$ & $\textbf{(423.10, 1782.14)}$ \\ [0.2ex]
Elimination &  $1169.41$ & $1063.32$ &  $926.46$ && $1330.14$ & $1140.98$ &   $1018.30$ && $1327.22$ & $1216.36$ &  $1076.24$ \\
 & $(644.26, 1712.07)$ & $(618.61, 1613.62)$ & $(582.42, 1475.14)$ && $(639.94, 1861.91)$ & $(584.48, 1702.59)$ & $(581.95, 1594.99)$ && $(593.47, 1887.17)$ & $(660.22, 1786.28)$ & $(605.23, 1669.82)$ \\ [0.2ex]
UCB & $899.96$ & $733.21$ &  $609.93$ && $1252.14$ & $1075.62$ & $921.17$ && $1545.74$ & $1379.55$ & $1222.05$ \\
 & $(850.05, 951.25)$ & $(691.01, 775.99)$ & $(574.80, 650.19)$ && $(1188.42, 1323.48)$ & $(1017.25, 1131.86)$ & $(868.39, 978.93)$ && $(1478.72, 1615.45)$ & $(1308.77, 1449.55)$ & $(1161.64, 1285.76)$ \\ [0.2ex]
Thompson & $1421.40$ & $1318.56$ &  $1226.48$ && $1844.29$ & $1783.12$ & $1692.16$ && $2008.64$ & $1970.39$ & $1919.84$ \\
 & $(677.00, 1818.72)$ & $(584.47, 1736.86)$ & $(545.91, 1651.00)$ && $(1418.01, 2036.27)$ & $(1363.05, 1994.79)$ & $(1035.50, 1931.36)$ && $(1837.35, 2111.36)$ & $(1763.34, 2078.33)$ & $(1691.93, 2055.71)$ \\ [0.2ex]
\bottomrule
\end{tabular}
}
\label{table:Beta-MARP}
\end{table}

Table \ref{table:Beta-MARP} reports the mean regret and its $90\%$ confidence interval under different  $m$ values and $c_t$ distributions for MARP, elimination, UCB and Thompson sampling. It is shown that MARP consistently exhibits the smallest mean regret among the algorithms.}

\subsection{Example on Bernoulli Rewards}
\label{sec:bernoullibandit}
\noindent
\change{The third example studies the case where  the stochastic reward 
\begin{equation}
\label{eqn:defbernoulibandit}
X_{i,t} = \sum_{j=1}^N \xi_j^{(i,t)}/N,
\end{equation} 
with $\xi_j^{(i,t)}$  drawn independently from $Bernoulli(\mu_i)$, $N=10^5$, $i = 1, 2, \cdots, m$ and $t = 1, 2, \cdots, T$. The mean reward $\mu_i$ drawn independently from $Beta(0.5, 3)$.   This reward structure follows the Bernoulli uplift bandits \citep{hsieh2022uplifting}.
Suppose there are $m=20$ arms. We consider two settings:  for ARP, the opportunity costs are public and uniform, $c_t = c_*$, $c_* \in \{0.05, 0.10, 0.15\}$; for MARP, $c_t$ follows a Beta distribution in $\{Beta(0.9, 0.9),$ $ Beta(1.1, 1.0),$ $ Beta(1.0, 1.1)\}$ unknown to the designer. 
To implement ARP, we set $k=10$, $\tau=0.2$ and compute $\theta_\tau$ in Eq.~\eqref{eqn:thetatau}, noting that $\min_{i\in\AA}\P(\mu_i-c_*\geq \tau) = 1 - F_x(c_* + \tau)$ where $F_x(\cdot)$ is the CDF of $Beta(0.5, 3)$. To ensure constraint Eq.~\eqref{eqn:publiceetinitial}, we set $\mu_1$ to $c_*$ in ARP. To implement the elimination algorithm, we use Algorithm 3 in \citep{even2006action} with $c=10$, $\delta=0.05$.
Parameters in the other methods can be calculated accordingly. The data are simulated by having $T = 15000$ agents for ARP and $T=8000$ agents for MARP.  We summarize results based on $500$ data realizations.

\begin{figure}[thb!]
\includegraphics[width=\textwidth]{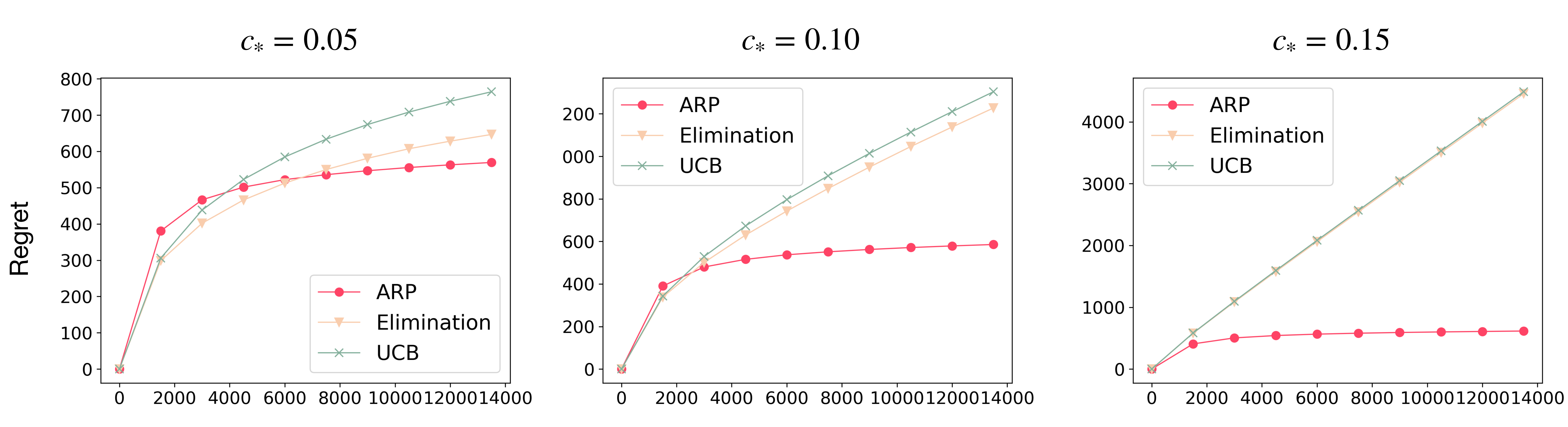}
\caption{The mean regret of ARP and alternative algorithms for Section \ref{sec:bernoullibandit}, based on 500 data replications. The three plots correspond to  $c_*=0.05$, $c_* = 0.10$, and $c_*=0.15$, respectively.}
\label{fig:simudata-ARP}
\end{figure}

Figure \ref{fig:simudata-ARP} presents the mean regret curve under different values of $c_*$ for ARP, elimination, and UCB. It is seen that  ARP outperforms the alternative algorithms across all scenarios. Moreover, the regret curve stabilizes after a sufficient number of rounds, indicating that ARP can identify the optimal arm. Initially, the regret curve for ARP is steep due to the sampling strategy of pulling all arms at a designated exploration rate while maintaining agents' incentives. This phase involves pulling suboptimal arms, which temporarily increases regret. However, ARP quickly transitions to selecting the optimal arm after the initial sampling stage.

\begin{figure}[thb!]
\includegraphics[width=\textwidth]{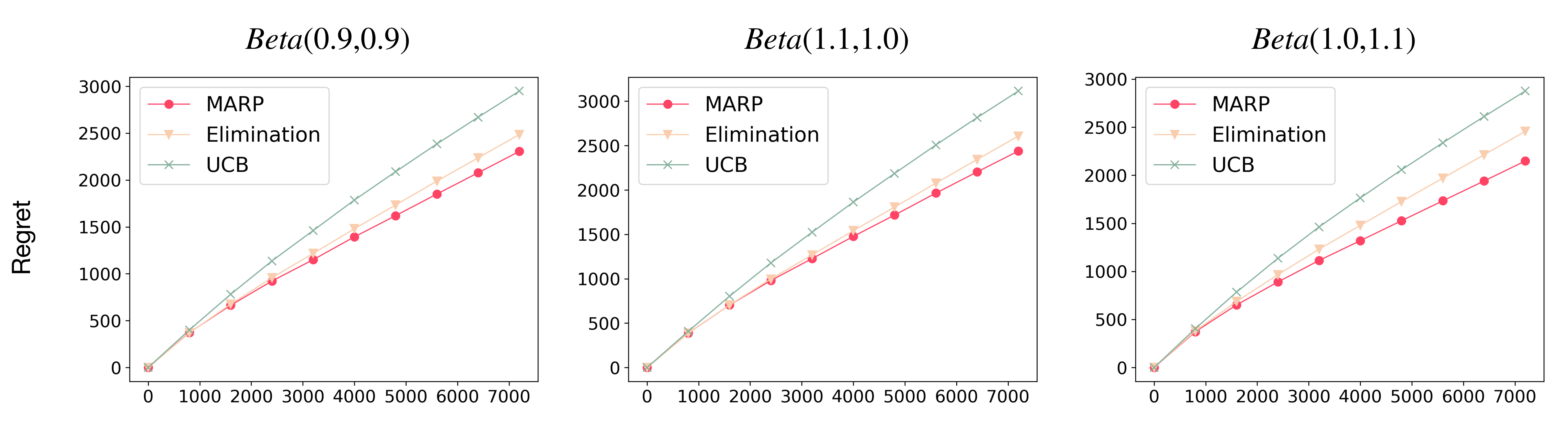}
\caption{The mean regret of MARP and alternative algorithms for Section \ref{sec:bernoullibandit}, based on 500 data replications. The three plots correspond to  $c_t\sim Beta(0.9, 0.9)$, $c_t\sim Beta(1.1, 1.0)$, and $c_t\sim Beta(1.0, 1.1)$, respectively.}
\label{fig:simudata-MARP}
\end{figure}

Figure \ref{fig:simudata-MARP} presents the mean regret curve under different $c_t$ distributions for MARP, elimination and UCB. It shows that MARP consistently exhibits the smallest mean regret among the algorithms due to its adaptive recommendation strategy based on individual loss rather than historical cumulative rewards.}

\section{Real Data Application}
\label{sec:realdataapplication}

\noindent
\change{In this section, we study a real data application using the Criteo Uplift Prediction Dataset \citep{diemert2018large}, which contains 25 million samples. Each sample represents an agent with 12 features, one binary exposure variable as treatment $D_j$, and one binary visit variable as outcome $Y_j$. We construct a Bernoulli uplifting bandit as Section \ref{sec:bernoullibandit}.
To build the model, we start by creating arm information. We randomly sample $10^5$ examples from the dataset without replacement and use $K$-means to partition these samples based on the features of agents, where $K=20$. We then assign the samples to $K$ clusters by calculating their K-means distance to the nearest cluster centroid. For cluster $i$, $i = 1, 2, \cdots, 20$, we calculate the expected reward $\mu_i$ for the corresponding arm according to $\mu_i = \sum_{j = 1}^{N_i} D_jY_j/ N_i$, where $N_i$ is the number of agents in the $i$-th cluster. We consider agents as homogeneous and arms as heterogeneous, where the number of arms equaling the number of clusters. 
Consider the Bernoulli reward  in Eq.~\eqref{eqn:defbernoulibandit} with $N=10^5$.
We compare ARP and MARP with alternative methods, including the upper confidence bound algorithm   (UCB) \citep{lai1985asymptotically} and the elimination algorithm \citep{even2006action}.

\begin{figure}[thb!]
\includegraphics[width=\textwidth]{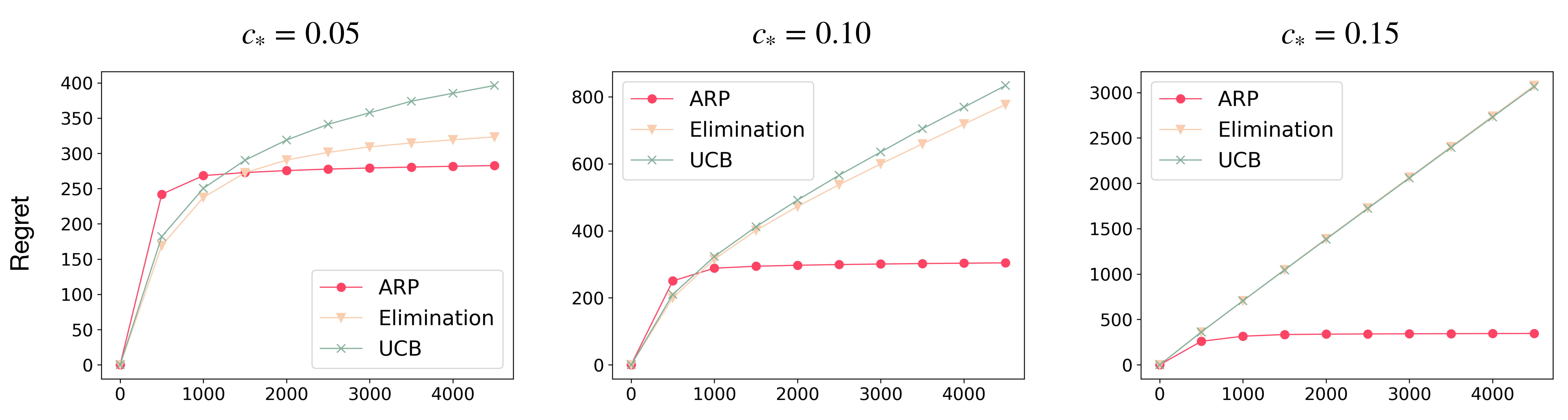}
\caption{The mean regret of ARP and alternative algorithms for Section \ref{sec:realdataapplication}, based on $200$ data replications. The three plots correspond to  $c_*=0.05$, $c_* = 0.10$, and $c_*=0.15$, respectively.}
\label{fig:realdata-ARP}
\end{figure}

In ARP, the opportunity costs are public and uniform, $c_t = c_*$, $c_*\in\{0.05, 0.10, 0.15\}$. We set $k=10$, $\tau=0.2$. The results are based on $T=5000$ agents and $200$ data replications.
Figure \ref{fig:realdata-ARP} demonstrates that ARP consistently outperforms alternative methods in terms of regret, particularly when the opportunity cost is high. This is because both UCB and the Elimination algorithm tend to fail when the historical average reward falls below the agent's opportunity costs, leading agents to disregard further recommendations and causing regret to skyrocket. In contrast, ARP successfully integrates the incentive constraint, ensuring that agents are motivated to follow the recommendations, thus maintaining lower levels of regret.
The performance of ARP is robust across different values of the opportunity cost $c_*$, which makes ARP highly effective in  online advertising and recommender systems where understanding of opportunity costs can be derived from prior research.

\begin{figure}[thb!]
\includegraphics[width=\textwidth]{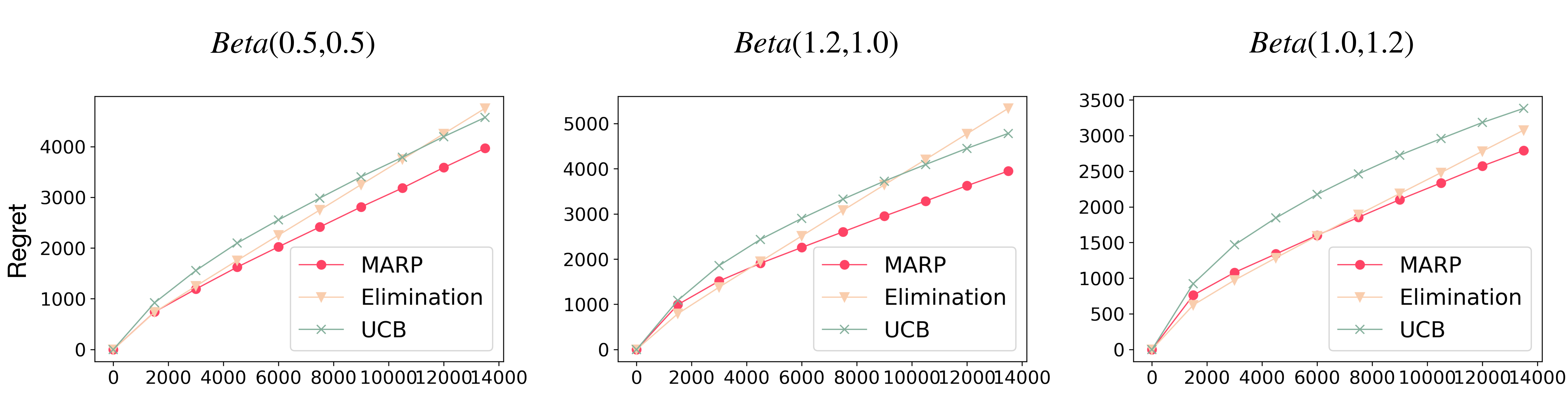}
\caption{The mean regret of MARP and alternative algorithms for Section \ref{sec:realdataapplication}, based on $200$ data replications. The three plots correspond to  $c_t\sim Beta(0.5, 0.5)$, $c_t\sim Beta(1.2, 1.0)$, and $c_t\sim Beta(1.0, 1.2)$, respectively.}
\label{fig:realdata-MARP}
\end{figure}

In MARP, the opportunity cost $c_t$ follows a Beta distribution  in $\{Beta(0.5, 0.5), Beta(1.2, 1.0), Beta(1.0, 1.2)\}$ unknown to the designer.  The results are based on $T=15000$ agents and $200$ data replications.
Figure \ref{fig:realdata-MARP} shows that MARP consistently achieves the lowest mean regret among the algorithms, across various settings. 
Particularly, we compare MARP with a competitive alternative, the elimination algorithm.  MARP dynamically adjusts recommendation probabilities and maintains exploration even after collecting extensive reward feedback. In contrast, the elimination algorithm tends to stop exploration prematurely after identifying a suboptimal choice, which leads to miss opportunities to identify the optimal arm and results in higher regret than MARP. The robust performance of MARP across different distributions of opportunity cost $\{c_t\}$ demonstrates its effectiveness in marketplaces where the agents' opportunity costs and their distributions are often undisclosed or confidential due to privacy concerns.}

\section{Discussion}
\label{sec:discussion}

\noindent
The design of recommender systems that respect incentives is a crucial need in order for users to be able to discover and choose valued products on a large scale. We have shown how information design and randomized recommendation can promote exploration, regardless of whether the agents' opportunity costs are known. A key aspect of an agent's incentive to explore is her belief about an arm. Our proposed recommendation policy allows the designer to pool recommendations across genuinely good arms and unknown arms. Although the agents in the latter case will never knowingly follow the recommendation, pooling the two circumstances for recommendations enables the designer to incentivize the agents to explore. 
We have also highlighted several factors that could improve exploration in online marketplaces, such as randomization, adaptivity, and background learning. \change{The code for reproducing the numerical results in this paper is available
at https://github.com/WenluValor/ARP-MARP.}

It is possible to extend the incentive-aware recommender system paradigm to other applications. For example, contextual bandits are important problems in machine learning, where each agent is characterized by a signal, called the context, observable by both the agent and the designer before the designer makes the recommendation. The context can include demographics, tastes, and other agent-specific information, and it impacts the expected rewards received by this agent. Contextual bandits are practically important: for instance, websites that make recommendations may possess a large amount of information about their users and would like to use this context to adjust their recommendations (e.g., Amazon and Netflix). 
Another intriguing question is to apply incentive-aware recommender systems to the adaptive clinical trial (ACT) designs for medical drugs. The ACT modifies the course of the trial based on the accumulating results of the trial, typically by adjusting the doses of medicine and adding or removing patients from the trial \citep{detry2012standards}.  Here an important aspect is the incentives for the patients and doctors to participate in and stay on the trial. It is crucial to manage their beliefs, which can be affected when the prescribed treatment changes throughout the trial.
We leave these questions for future research.

\begin{acks}
The authors thank three anonymous reviewers, the Associate Editor, and the Editor for their invaluable feedback. 
\end{acks}

\bibliographystyle{ACM-Reference-Format}
\bibliography{match}

\newpage
\appendix

\section{Technical Proofs}

\subsection{Proof of Theorem  \ref{thm:knowncost}}

\begin{proof}
ARP has two main steps: (i) the \emph{sampling step}, which collects $k$ rewards from each arm, and (ii) the \emph{exploration-exploitation step}, where the agents explore the arms based on the collected sample and then identify an exploit arm. We now discuss the incentive constraints for the two steps separately.

\paragraph{The sampling step} This phase can be divided into $m$ stages, where each stage, $i=1,\ldots,m$, lasts $L_i$ rounds and $L_1=1$. We will show that each agent $t$ in each stage $i$ is incentivized to pull the recommended arm. For an agent $t$ in the stage $i$, the incentive constraint in Eq.~\eqref{eqn:incentivesct} is equivalent to
\begin{equation}
\label{eqn:incentivesampling}
\E[X_{I_t,t}-c_*|I_t=i,\EE_t]\cdot \P(I_t=i)\geq 0.
\end{equation}
Agent $t$ is in the ``explore group" with probability $p_{i,t}$ and in the ``exploit group"  with probability $1-p_{i,t}$. \change{Then,
\begin{equation*}
\begin{aligned}
\E[X_{I_t,t}-c_*|I_t=i,\EE_t]\cdot \P(I_t=i) & =p_{i,t}\cdot\E[X_{I_t,t}-c_*|I_t=i,\EE_t,\text{``explore group"}]\cdot \P(I_t=i)  \\
& \quad + (1-p_{i,t})\cdot\E[X_{I_t,t}-c_*|I_t=i,\EE_t,\text{``exploit group"}]\cdot \P(I_t=i). 
\end{aligned}
\end{equation*}}
Conditional on being in the explore group and seeing the public information $\EE_t$ in Eq.~\eqref{eqn:publiceet}, the expected gain from following the recommendation $I_t$ is
 $\widehat{M}_i$. Here $\widehat{M}_i$ is the empirical mean of historical rewards up to stage $i-1$,
\begin{equation*}
\widehat{M}_i = \frac{1}{\sum_{j=1}^{i-1}L_j}\sum_{s=1}^{L_1+\cdots+L_{i-1}}X_{I_s,s} y_s.
\end{equation*}
 On the other hand, conditional on being in the exploit group and conditional on $\EE_t$, the expected gain from following the recommendation $I_t$ can be analyzed as follows.
Define the event $\CC_1$ as,
\begin{equation*}
\CC_1 = \left\{\hat{\mu}_{a_i^*}^k-\lambda\geq c_*\right\}.
\end{equation*} 
Then the expected gain from following the recommendation $I_t$ is at least 
\begin{equation*}
\E[X_{I_t,t}|I_t=a_i^*,\EE_t,\CC_1]\cdot\P(\CC_1).
\end{equation*} 
Combining the above two scenarios, we have
\begin{equation*}
\begin{aligned}
&\E[X_{I_t,t}-c_*|I_t=i,\EE_t]\cdot \P(I_t=i) \\
& \geq  \left(\widehat{M}_i-c_*\right)p_{i,t} +  \E[X_{I_t,t}-c_*|I_t=a_i^*,\EE_t,\CC_1]\P(\CC_1)\left(1-p_{i,t}\right).
\end{aligned}
\end{equation*}
Thus,  it suffices to choose $p_{i,t}$  as follows  to guarantee the constraint in Eq.~\eqref{eqn:incentivesampling},
\begin{equation*}
p_{i,t}\leq \frac{\E[X_{I_t,t}-c_*|I_t=a_i^*,\EE_t,\CC_1]\P(\CC_1)}{\E[X_{I_t,t}-c_*|I_t=a_i^*,\EE_t,\CC_1]\P(\CC_1) + c_*-\widehat{M}_i}.
\end{equation*}
Define the parameter $\epsilon \equiv \lambda\sigma_i/3$. 
Recall that $\hat{\mu}^k_i$ is the empirical mean reward of arm $i$.
Consider the event
\begin{equation*}
\CC_2 = \{\forall i\in\AA: |\hat{\mu}^k_i-\mu_i|\leq \epsilon\}.
\end{equation*}
Using the Chernoff-Hoeffding bound and the union bound, 
\begin{equation*}
\forall \mu_1,\ldots,\mu_m:\ \P(\CC_2)\geq 1-2m\cdot e^{-2\epsilon^2k}.
\end{equation*}
From  Eq.~\eqref{eqn:positiveplat}, $\mu_1>c_*$. We  let  $0<\lambda\leq 3(\mu_1-c_*)/4$. Then under the event $\CC_2$, 
\begin{equation*}
\begin{aligned}
\hat{\mu}_1^k-\lambda\sigma_1 & = (\hat{\mu}_1^k- \mu_1) + (\mu_1 -\lambda\sigma_1)\\
& \geq -\epsilon + (\epsilon+c_*) = c_*.
\end{aligned}
\end{equation*}
By definition, $\hat{\mu}_{a_i^*}^k-\lambda\geq \hat{\mu}_1^k-\lambda$. Hence we have 
\begin{equation*}
\P(\CC_1|\CC_2)=1 \text{ and } \P(\CC_1,\CC_2) = \P(\CC_2)\P(\CC_1|\CC_2) = \P(\CC_2),
\end{equation*}
which implies that 
\begin{equation*}
\begin{aligned}
&\E[X_{I_t,t}-c_*|I_t=a_i^*,\EE_t,\CC_1]\P(\CC_1) \\
& = \E[X_{I_t,t}-c_*|I_t=a_i^*,\EE_t,\CC_1,\CC_2]\P(\CC_1,\CC_2)+\E[X_{I_t,t}-c_*|I_t=a_i^*,\EE_t,\CC_1,\neg\CC_2]\P(\neg\CC_2)\\
&\geq \E[X_{I_t,t}-c_*|I_t=a_i^*,\EE_t,\CC_1,\CC_2]\P(\CC_2)-2m\cdot e^{-2\epsilon^2k}.
\end{aligned}
\end{equation*}
Observe that, 
\begin{equation*}
\begin{aligned}
 & \E[X_{I_t,t}-c_*|I_t=a_i^*,\EE_t,\CC_1,\CC_2] \\
 & =  \E[X_{I_t,t}-\hat{\mu}_i^k|I_t=a_i^*,\EE_t,\CC_1,\CC_2] + \E[\hat{\mu}_i^k-c_*|I_t=a_i^*,\EE_t,\CC_1,\CC_2] \\
&\geq -\epsilon + \lambda. 
\end{aligned}
\end{equation*}
Thus, 
\begin{equation*}
\begin{aligned}
& \E[X_{I_t,t}-c_*|I_t=a_i^*,\EE_t,\CC_1]\P(\CC_1) \\
& \geq (\lambda-\epsilon)\P(\CC_2)-2m\cdot e^{-2\epsilon^2k}\\
& \geq (\lambda-\epsilon)\left(1-2m\cdot e^{-2\epsilon^2k}\right) - 2m\cdot e^{-2\epsilon^2k}.
\end{aligned}
\end{equation*}
We let
\begin{equation*}
k\geq \frac{9}{2\lambda^2}\ln\left(\frac{20m}{\lambda}\right).
\end{equation*} 
Then,
\begin{equation*}
\begin{aligned}
\E[X_{I_t,t}-c_*|I_t=a_i^*,\EE_t,\CC_1]\P(\CC_1) \geq \frac{3}{5}\lambda-\frac{1}{10}\lambda = \frac{1}{2}\lambda.
\end{aligned}
\end{equation*}
Thus, it suffices to choose
\begin{equation*}
p_{i,t}\leq \frac{\lambda}{2(c_*-\widehat{M}_i)+\lambda}.
\end{equation*}
Since $0\leq p_{i,t}\leq 1$, it suffices to choose 
\begin{equation*}
p_{i,t} = \frac{\lambda}{2(c_*-\widehat{M}_i)+\lambda}\mathbf{1}\{0\leq \widehat{M}_i< c_*\} + \mathbf{1}\{\widehat{M}_i\geq c_*\}.
\end{equation*}

\paragraph{The exploration-exploitation step}  We will show that any agent $t$ in this phase satisfies the incentive constraint in  Eq.~\eqref{eqn:incentivesampling},
\begin{equation*}
\E[X_{I_t,t}-c_*|I_t=i,\EE_t]\cdot \P(I_t=i)\geq 0.
\end{equation*}
Let $l_*=\theta_\tau^2\ln (T\theta_\tau)$. By Eq.~\eqref{eqn:choiceofk}, $k\geq l_*$. Let $\gamma_l=\sqrt{\frac{\ln(T\theta_\tau)}{2l}}$ for any $l\geq k$. We have that
\begin{equation}
\label{eqn:gammantau4}
\begin{aligned}
\gamma_l&\leq\gamma_{l_*} = \sqrt{\frac{\ln(T\theta_\tau)}{2l_*}} =  \sqrt{\frac{\ln(T\theta_\tau)}{2\theta_\tau^2\ln (T\theta_\tau)}} 
= \frac{\tau\cdot\min_{i\in\AA}\P(\mu_i-c_*\geq \tau)}{4\sqrt{2}m^2} \\
& <  \frac{\tau}{4}\min_{i\in\AA}\P(\mu_i-c_*\geq \tau),\quad\forall l\geq k.
\end{aligned}
\end{equation}
Define the following event,
\begin{equation*}
\CC_3=\left\{\forall l\geq k,\forall i\in\AA:|\hat{\mu}_{i}^l - \mu_{i}|<\gamma_l\right\}.
\end{equation*}
By the Chernoff-Hoeffding bound and the union bound, we have
\begin{equation*}
\begin{aligned}
\P(\neg\CC_3) & \leq\sum_{l\geq k}\sum_{i\in\AA}\P(|\hat{\mu}_{i}^l-\mu_{i}|\geq \gamma_l)\leq\sum_{l\geq k}\sum_{i\in\AA}e^{-2l\cdot \gamma_l^2}\\
& \leq T\cdot m\cdot e^{-\ln (T\theta_\tau)}\leq\frac{m}{\theta_\tau}.
\end{aligned}
\end{equation*}
Now we analyze the incentive condition. 
Since $X_{I_t,t}\in[0,1]$ and $c_*\in(0,1)$, we have
\begin{equation}
\label{eqn:bdoneziip}
\begin{aligned}
& \E[X_{I_t,t}-c_*|I_t=i,\EE_t]\cdot\P(I_t=i)\\
&=\E[X_{I_t,t}-c_*|I_t=i,\EE_t,\CC_3]\cdot\P(I_t=i,\CC_3)\\
&\quad\quad+\E[X_{I_t,t}-c_*|I_t=i,\EE_t,\neg\CC_3]\cdot\P(I_t=i,\neg\CC_3)\\
&\geq \E[X_{I_t,t}-c_*|I_t=i,\EE_t,\CC_3]\cdot\P(I_t=i,\CC_3)-\frac{m}{\theta_\tau}.
\end{aligned}
\end{equation}
We shall focus on the integral $\E[X_{I_t,t}-c_*|I_t=i,\EE_t,\CC_3]$.
If  $\E[X_{I_t,t}-c_*|I_t=i,\EE_t,\CC_3]\leq -2\gamma_l$, then because 
\begin{equation*}
\hat{\mu}_i^l - c_* < \mu_i - c_* + \gamma_l \leq -\gamma_l,
\end{equation*}
we must have already eliminated this arm $i$.
Thus, in that case, $I_t=i$ cannot occur. If $\E[X_{I_t,t}-c_*|I_t=i,\EE_t,\CC_3]\geq -2\gamma_l$ and have the following lower bound,
\begin{equation*}
\begin{aligned}
& \E[X_{I_t,t}-c_*|I_t=i,\EE_t,\CC_3]\cdot\P(I_t=i,\CC_3)\\
&\geq \tau\cdot\P(\CC_3,\mu_i-c_*\geq \tau) + 0\cdot\P(\CC_3,0\leq \mu_i - c_*\leq \tau) \\
&\quad\quad -2 \gamma_l\cdot\P(\CC_3,-2\gamma_l\leq \mu_i-c_*\leq 0)\\
&\geq \tau\cdot\P(\CC_3,\mu_i-c_*\geq \tau)-2 \gamma_l\\
&> \tau\cdot\P(\CC_3, \mu_i-c_*\geq\tau)-\frac{\tau\cdot\P(\mu_i-c_*\geq \tau)}{2},
\end{aligned}
\end{equation*}
where the last step is by Eq.~\eqref{eqn:gammantau4}.
Moreover, we can lower bound $\P(\CC_3,\mu_i-c_*\geq \tau)$ using the Chernoff-Hoeffding bound,
\begin{equation*}
\begin{aligned}
\P(\CC_3,\mu_i-c_*\geq \tau) & = \P(\CC_3|\mu_i-c_*\geq \tau)\cdot\P(\mu_i-c_*\geq \tau)\\
&\geq\left(1-\sum_{l\geq k}\sum_{i\in\AA}e^{-2l\cdot\gamma_l^2}\right)\cdot\P(\mu_i-c_*\geq \tau)\\
&\geq\left(1-\frac{m}{\theta_\tau}\right)\cdot\P(\mu_i-c_*\geq\tau)\\
&>\frac{3}{4}\P(\mu_i-c_*\geq\tau),
\end{aligned}
\end{equation*}
where last step above is due to $\theta_\tau = 4m^2/[\tau\min_{i\in\AA}\P(\mu_i-c_*\geq \tau)]$, and
\begin{equation}
\frac{m}{\theta_\tau} = \frac{m}{4m^2}\cdot\tau\cdot \min_{i\in\AA}\P(\mu_i-c_*\geq \tau)<\frac{1}{4}.
\end{equation}
Thus, we have
\begin{equation*}
\E[X_{I_t,t}-c_*|I_t=i,\CC_3]\cdot\P(I_t=i,\EE_t,\CC_3)>\frac{\tau}{4}\cdot\P(\mu_i-c_*\geq\tau).
\end{equation*}
Plugging this back to Eq.~\eqref{eqn:bdoneziip} and using the fact that $\theta\geq\theta_\tau$, we obtain that
\begin{equation*}
\begin{aligned}
& \E[X_{I_t,t}-c_*|I_t=i,\EE_t]\cdot \P(I_t=i)\\
& \geq \E[X_{I_t,t}-c_*|I_t=i,\EE_t,\CC_3]\P(I_t=i,\CC_3)-\frac{m}{\theta_\tau}\\
&>\frac{\tau}{4} \cdot\P(\mu_i-c_*\geq\tau) - \frac{m}{\theta_\tau}\\
& \geq \frac{\tau}{4} \cdot\P(\mu_i-c_*\geq\tau) -\frac{\tau}{4m} \cdot\min_{i\in\AA}\P(\mu_i-c_*\geq\tau)\\
&\geq 0,
\end{aligned}
\end{equation*}
which proves the incentive constraint in Eq.~\eqref{eqn:incentivesampling}.
\end{proof} 

\subsection{Proof of Theorem \ref{lem:individualfair}}
\begin{proof}
Let the criterion $\HH(t)$ be defined in Eq.~\eqref{eqn:egofht} for any $t\geq 1$, where each agent has a context $(\alpha_t,\beta_t)$,  where $\alpha_t$ is the total number of agent $t$'s visits up to (but not including) round $t$. The $\beta_t$ is the number of visits up to (but not including) round $t$, where she followed the designer's recommendation but learned that the received reward was worse than $c_t$.

For any $t\geq 1$, we consider two cases. First, if the agent $t$ is new to the market, we have $\alpha_t=\beta_t=0$. According to the recommendation policy in Eq.~\eqref{eqn:randrecpolicyindfair}, if $\HH(t)$ holds, that is, 
\begin{equation*}
    \frac{\beta_t+1}{\alpha_t+1}\leq \gamma_{t},
\end{equation*}
then $I_t\in\{a_i^*,i\}$, and 
\begin{equation*}
    \frac{\beta_t+\mathbf{1}\{\E[X_{I_t,t}]<c_t\}}{\alpha_t+1}\leq \frac{\beta_t+1}{\alpha_t+1}\leq \gamma_{t}.
\end{equation*}
If  $\HH(t)$ does not hold, that is, 
\begin{equation*}
    \frac{\beta_t+1}{\alpha_t+1}> \gamma_{t},
\end{equation*}
then $I_t=a_i^*$. By assumption \eqref{eqn:positiveplat}, $\E[X_{I_t,t}]\geq \mu_1> c_t$. Thus,
\begin{equation*}
    \frac{\beta_t+\mathbf{1}\{\E[X_{I_t,t}]<c_t\}}{\alpha_t+1}= \frac{\beta_t}{\alpha_t+1}=0\leq \gamma_{t}.
\end{equation*}
By definition, the ARP procedure in this case satisfies ex-post fairness in Eq.~\eqref{def:expostfair}.

Second, if the agent $t$ is a returned agent to the market, we have $\alpha_t\neq 0$ and 
\begin{equation}
\label{eqn:returncond}
    \frac{\beta_t}{\alpha_t}\leq \gamma_t.
\end{equation} Similarly, according to the recommendation policy in Eq.~\eqref{eqn:randrecpolicyindfair}, if $\HH(t)$ holds, the proof follows from the first case.
If  $\HH(t)$ does not hold,
then $I_t=a_i^*$. By assumption \eqref{eqn:positiveplat}, $\E[X_{I_t,t}]\geq \mu_1> c_t$. Thus,
\begin{equation*}
    \frac{\beta_t+\mathbf{1}\{\E[X_{I_t,t}]<c_t\}}{\alpha_t+1}= \frac{\beta_t}{\alpha_t+1}\leq\frac{\beta_t}{\alpha_t}\leq \gamma_{t},
\end{equation*}
where the last step is by Eq.~\eqref{eqn:returncond}.
By definition, the ARP procedure  in this case satisfies ex-post fairness in Eq.~\eqref{def:expostfair}.
This concludes the proof. 
\end{proof}

\subsection{Proof of Theorem \ref{lem:regretofknowncost}}
\begin{proof}
We divide the proof into two parts: the regret from the \emph{incentive} part (i.e., Step 2 of ARP), and the regret from the \emph{exploration-exploitation} part (i.e., Steps 3-4 of ARP).
For the incentive part, denote the gap of rewards as,
\begin{equation*}
\Delta=\mu^*-\max_{i:\mu_i<\mu^*}\mu_i\leq 1,\quad\text{where}\quad\mu^*=\max_{i\in\AA}\mu_i.
\end{equation*}
The gap is the difference between the largest and the second largest expected reward.
Since the parameters $\lambda,\theta_\tau, k$ are chosen according to Eqs.\ \eqref{eqn:choiceoflambda}, \eqref{eqn:thetatau}, and \eqref{eqn:choiceofk}, Theorem \ref{thm:knowncost} guarantees the agent's incentive. Moreover, given $\gamma_t=1$ in Eq.~\eqref{eqn:egofht}, the criterion $\HH(t)$ defined in \eqref{eqn:egofht} is satisfied. Thus, ex-post regret is upper bounded by,
\begin{equation}
\label{eqn:incentiveregret}
L^*\Delta 
\end{equation}
For the exploration-exploitation part, it is shown in \citet{even2006action} and Lemma 7 in \citet{mansour2020bayesian} that there exists a logarithmic bound for ex-post regret,
\begin{equation*}
\sum_{i\in\AA}\frac{18\log (T\theta_\tau)}{\mu^*-\mu_i}.
\end{equation*}
By definition of $\Delta$, the above result can be further bounded by,
\begin{equation}
\label{eqn:explorebound}
\frac{18m\log (T\theta_\tau)}{\Delta}.
\end{equation}
Finally, combining Eqs.~\eqref{eqn:incentiveregret} and \eqref{eqn:explorebound} yields that,
\begin{equation*}
R_T \leq L^*\Delta + \frac{18m\log (T\theta_\tau)}{\Delta}.
\end{equation*}
On the other hand, $R_T$ can be upper bounded by $\Delta$ per round. Hence,
\begin{equation*}
R_T \leq \min\left\{L^*\Delta + \frac{18m\log (T\theta_\tau)}{\Delta},T\Delta\right\}.
\end{equation*}
By the Cauchy-Schwarz inequality and the fact that $\Delta\leq 1$, we have
\begin{equation*}
R_T \leq L^* + O\left(\sqrt{mT\cdot\log(T\theta_\tau)}\right).
\end{equation*}
By definition in Eq.~\eqref{eqn:thetatau}, $\theta_\tau=O(m)$.
This completes the proof.
\end{proof}

\subsection{Proof of Theorem \ref{lem:individualfairmarp2}}
\begin{proof}
Note that the proof follows along exactly the same lines as the proof of Theorem \ref{lem:individualfair} and thus we omit the details. 
\end{proof}

\subsection{Proof of Theorem \ref{thm:regretexpweight}}
\begin{proof}
First, we define an expected loss $\bar{l}$ as follows,
\begin{equation}
\label{eqn:defofbarl}
\bar{l}(\bp_t,y_t) \equiv \E_{I_t}[\hat{l}(I_t,y_t)] = \sum_{i=1}^m \hat{l}(i,y_t)p_{i,t},
\end{equation}
where $\hat{l}(\cdot,\cdot)$ is defined in \eqref{eqn:estimatedloss}.
Since $\hat{l}\in[0,1]$, we have $\bar{l}\in[0,1]$.
Define an instantaneous regret vector, $\br_t\equiv (r_{1,t},\ldots,r_{m,t})\in\R^m$, with the elements defined as
$r_{i,t} \equiv \bar{l}(\bp_t,y_t)-\hat{l}(i,y_t)$.
Then $r_{i,t} $ measures the expected change in the designer's loss if it were to recommend arm $i$ and agent $t$ did not change action $y_t$.  Write the regret vector $\bar{\br}_t \equiv (\bar{r}_{1,t},\ldots,\bar{r}_{m,t})\in\R^m$, which is defined by
\begin{equation*}
\bar{\br}_t=\sum_{s=1}^t\br_s,\quad \forall t\geq 1.
\end{equation*} 
Let $\Phi(\cdot)$ be a potential function $\Phi:\R^m\to\R$, defined as,
\begin{equation*}
\Phi(\bar{\br}_t) = \frac{1}{\eta}\ln\left(\sum_{i=1}^m\exp(\eta\cdot \bar{r}_{i,t})\right),\quad \forall t\geq 1\text{ and }i=1,\ldots,m.
\end{equation*}
The probability $p_{i,t}$ in Eq.~\eqref{eqn:expweightstrategy} is equivalent to
\begin{equation*}
p_{i,t}= \frac{\partial_i\Phi(\bar{\br}_{t-1})}{\sum_{j=1}^m\partial_j\Phi(\bar{\br}_{t-1})},
\end{equation*}
for $t>1$ and $p_{i,1}=1/m$, where $i=1,\ldots,m$. 
Since  $\bar{l}$ in Eq.~\eqref{eqn:defofbarl} is linear in $\bp_t$, then  
\begin{equation*}
\begin{aligned}
\bar{l}(\bp_t,y_t) = \bar{l}\left(\left\{\frac{\partial_i\Phi(\bar{\br}_{t-1})}{\sum_{j=1}^m\partial_j\Phi(\bar{\br}_{t-1})}\right\}_{i=1}^m,y_t\right) = \frac{\sum_{i=1}^m\partial_i\Phi(\bar{\br}_{t-1})\hat{l}(i,y_t)}{\sum_{j=1}^m\partial_j\Phi(\bar{\br}_{t-1})}.
\end{aligned}
\end{equation*}
Rearranging, we obtain that for any $y_t\in\YY$,
\begin{equation}
\label{eqn:blackwellcond}
\sum_{i=1}^mp_{i,t}\left[\bar{l}(\bp_{t},y_t)-\hat{l}(i,y_t)\right] = 0.
\end{equation}
Note that the condition in Eq.~\eqref{eqn:blackwellcond} is similar to a key property used in the proof of Blackwell's celebrated approachability theorem~\citep{blackwell1956analog}.

Next, we want to upper bound a regret $\bar{R}_T\in\R$ defined by
\begin{equation}
\label{eqn:defofbarRT}
\bar{R}_T \equiv \sum_{t=1}^T\bar{l}(\bp_t,y_t) - \min_{i\in\AA}\sum_{t=1}^T\hat{l}(i,y_t).
\end{equation}
It is seen that $\bar{R}_T$ measures the difference between the cumulative expected loss and the loss of the optimal arm.
We show that for any $T\geq 1$ and $y_1,\ldots,y_T\in\YY$,  
\begin{equation}
\label{eqn:them2.2inthebook}
\bar{R}_T\leq   \sqrt{\frac{T}{2}\ln m}.
\end{equation} 
To prove Eq.~\eqref{eqn:them2.2inthebook}, we introduce some additional notation. Let
\begin{equation*}
W_t \equiv \sum_{i=1}^m\exp\left(-\eta\cdot \sum_{s=1}^t\hat{l}(i,y_s)\right) \text{ for } t\geq 1,
\end{equation*}
and let $W_0=m$. Observe that
\begin{equation}
\label{eqn:lowbdwnw0}
\begin{aligned}
\ln\frac{W_T}{W_0}   &\geq \ln\left(\max_{i\in\AA}\exp\left(-\eta\cdot \sum_{s=1}^T\hat{l}(i,y_s)\right)\right) - \ln m\\
& = -\eta\cdot\min_{i\in\AA} \sum_{s=1}^T\hat{l}(i,y_s) - \ln m.
\end{aligned}
\end{equation}
On the other hand, for each $t=1,\ldots,T$,
\begin{equation*}
\begin{aligned}
\ln\frac{W_{t}}{W_{t-1}} & = \ln\frac{\sum_{i=1}^m\exp\left(-\eta\cdot \hat{l}(i,y_{t})\right)\cdot\exp\left(-\eta\cdot \sum_{s=1}^{t-1}\hat{l}(i,y_s)\right)}{\sum_{j=1}^m\exp\left(-\eta\cdot \sum_{s=1}^{t-1}\hat{l}(j,y_s)\right)}\\
& = \ln\frac{\sum_{i=1}^mp_{i,t} \cdot\exp\left(-\eta\cdot \hat{l}(i,y_{t})\right)}{\sum_{j=1}^mp_{j,t}}.
\end{aligned}
\end{equation*}
Since $\hat{l}\in[0,1]$, Hoeffding's lemma  \citep{massart2007concentration} yields
\begin{equation*}
\begin{aligned}
\ln\frac{\sum_{i=1}^mp_{i,t}\cdot \exp\left(-\eta\cdot \hat{l}(i,y_{t})\right)}{\sum_{j=1}^mp_{j,t}} & \leq  -\eta\cdot \frac{\sum_{i=1}^mp_{i,t}\cdot \hat{l}(i,y_t)}{\sum_{j=1}^mp_{j,t}}+\frac{\eta^2}{8} \\
& =-\eta \cdot \bar{l}(\bp_{t},y_t)+\frac{\eta^2}{8},
\end{aligned}
\end{equation*}
where we used Eq.~\eqref{eqn:blackwellcond} in the last step. Summing over $t=1,\ldots,T$, we get
\begin{equation*}
\ln\frac{W_T}{W_0} \leq -\eta\cdot\sum_{t=1}^T\bar{l}(\bp_{t},y_t)+\frac{\eta^2}{8}T.
\end{equation*}
Combining this with the lower bound in Eq.~\eqref{eqn:lowbdwnw0}, we find that
\begin{equation*}
\sum_{t=1}^T\bar{l}(\bp_{t},y_t)\leq\min_{i\in\AA} \sum_{t=1}^T \hat{l}(i,y_t) + \frac{\ln m}{\eta} + \frac{\eta\cdot T}{8}.
\end{equation*}
Letting $\eta=\sqrt{(8\ln m)/T}$,  the above inequality yields Eq.~\eqref{eqn:them2.2inthebook}.
Taking the expectations on both sides of Eq.~\eqref{eqn:them2.2inthebook}, we obtain that
\begin{equation}
\label{eqn:them2.2inthebookexp}
 \sum_{t=1}^T\E[\bar{l}(\bp_t,y_t)] - \min_{i\in\AA}\sum_{t=1}^Tl(i,y_t) \leq   \sqrt{\frac{T}{2}\ln m}.
\end{equation}

Finally, we note that the random variables $\{l(I_t,y_t)-\E[\bar{l}(\bp_t,y_t)],t\geq 1\}$ form a sequence of bounded martingale differences. By the Hoeffding-Azuma inequality, with probability at least $1-\delta$ for any $\delta\in(0,1)$, 
\begin{equation}
\label{eqn:bdonhalIp}
\sum_{t=1}^Tl(I_t,y_t)\leq\sum_{t=1}^T\E[\bar{l}(\bp_t,y_t)]+\sqrt{\frac{T}{2}\ln\frac{1}{\delta}}.
\end{equation}
We complete the proof by combining this inequality with Eq.~\eqref{eqn:them2.2inthebookexp}.
\end{proof}

\subsection{Proof of Theorem \ref{thm:lowerbd}}

\begin{proof}
Note that the minimax regret satisfies 
\begin{equation}
\label{eqn:boundoninfsup}
\inf_{\{I_t\}_{t=1}^T}\sup_{\{\AA:|\AA|=m\}, \{y_t\}_{t=1}^T\in\YY^T} R_T \geq \sup_{\{\AA:|\AA|=m\}} \inf_{\{I_t\}_{t=1}^T}\sup_{\{y_t\}_{t=1}^T\in\YY^T}R_T.
\end{equation}

First, we consider a fixed arm set $\AA$ and give a lower bound for
$\inf_{\{I_t\}_{t=1}^T}\sup_{\{y_t\}_{t=1}^T\in\YY^T}R_T$.
We consider that $\mu_i$'s are i.i.d. random variables with  
\begin{equation*}
\P[\mu_i=0]=\P[\mu_i=1]=\frac{1}{2}.
\end{equation*}
Let the actions $\tilde{y}_1,\ldots,\tilde{y}_T$ be i.i.d. random variables with 
\begin{equation*}
\P[\tilde{y}_t=0] = \epsilon_T, \P[\tilde{y}_t=1]=1-\epsilon_T,
\end{equation*} where $\epsilon_T>0$ is any parameter satisfying $\lim_{T\to\infty}\epsilon_{T}=0$. Then 
\begin{equation*}
\begin{aligned}
\inf_{\{I_t\}_{t=1}^T}\sup_{\{y_t\}_{t=1}^T\in\YY^T}R_T & \geq \inf_{\{I_t\}_{t=1}^T}\E\left\{\sum_{t=1}^Tl(I_t, \tilde{y}_t)-\min_{i\in\AA}\sum_{t=1}^Tl(i, \tilde{y}_t)\right\}\\
& = \inf_{\{I_t\}_{t=1}^T}\E\sum_{t=1}^Tl(I_t,\tilde{y}_t) -\E\min_{i\in\AA}\sum_{t=1}^Tl(i,\tilde{y}_t).
\end{aligned}
\end{equation*}
Since $\{\tilde{y}_1,\ldots, \tilde{y}_T\}$ are completely random, 
\begin{equation*}
\begin{aligned}
& \inf_{\{I_t\}_{t=1}^T}\sup_{\{y_t\}_{t=1}^T\in\YY^T}R_T \\
& \geq \left[ \epsilon_TT + (1-\epsilon_T)\E\sum_{t=1}^T\left(1-\mu_{I_t}\right)  \right] - \E\min_{i\in\AA}\left[ \epsilon_TT + (1-\epsilon_T)\sum_{t=1}^T\left(1-\mu_{i,t}\right) \right]\\
& =  \left[\epsilon_T T + \frac{1}{2}(1-\epsilon_T)T \right] -  \E\min_{i\in\AA}\left[ \epsilon_TT + \frac{1}{2}(1-\epsilon_T)T  + (1-\epsilon_T)\sum_{t=1}^T\left(\frac{1}{2}-\mu_{i,t}\right)  \right]\\
& =  \frac{1}{2}(1-\epsilon_T)\E\left[\max_{i=1,\ldots,m}\sum_{t=1}^TZ_{i,t}\right],
\end{aligned}
\end{equation*}
where each $Z_{i,t}$ is an i.i.d.\ Rademacher random variable:
\begin{equation*}
\P[Z_{i,t}=-1]=\P[Z_{i,t}=1]=1/2.
\end{equation*} 
Then
\begin{equation}
\label{eqn:lowerbdonrt}
\begin{aligned}
\sup_{\{\AA:|\AA|=m\}}\inf_{\{I_t\}_{t=1}^T}\sup_{\{y_t\}_{t=1}^T\in\YY^T}R_T \geq \frac{1}{2}(1-\epsilon_T)\E\left[\max_{i=1,\ldots,m}\sum_{t=1}^TZ_{i,t}\right].
\end{aligned}
\end{equation}

Second, we define a $m$-vector $\tilde{\mathbf Z}_T=(\tilde{Z}_{T,1},\ldots,\tilde{Z}_{T,m})$ with entries,
\begin{equation}
\label{eqn:defoftildez}
\tilde{Z}_{T,i}\equiv \frac{1}{\sqrt{T}}\sum_{t=1}^TZ_{i,t},\quad i=1,\ldots,m.
\end{equation}
By the central limit theorem, for any vector  $(a_1,\ldots,a_m)\in\R^m$, the summation $\sum_{i=1}^ma_i\tilde{Z}_{T,i}$ converges in distribution, as $T\to\infty$, to a zero-mean normal random variable with variance $\sum_{i=1}^ma_i^2$. By the  Cramér–Wold theorem, the vector $\tilde{\mathbf Z}_T$ converges in distribution to $\mathbf G=(G_1,\ldots,G_m)$, where each $G_i$ is an i.i.d.\ standard normal random variable.
Then for any bounded continuous function $\psi:\R^m\to\R$,
\begin{equation}
\label{eqn:limTpsi}
\lim_{T\to\infty}\E\left[\psi(\tilde{Z}_{T,1},\ldots,\tilde{Z}_{T,m})\right] = \E\left[\psi(G_1,\ldots,G_m)\right].
\end{equation}
Denote  a parameter $\tilde{L}>0$. Define a function $\phi_{\tilde{L}}$ as,
\begin{equation*}
\phi_{\tilde{L}}(z) = -\tilde{L}\mathbf{1}(z<-\tilde{L}) + z \mathbf{1}(|z|\leq \tilde{L}) + \tilde{L}\mathbf{1}(z>\tilde{L}),
\end{equation*}
and let $\psi(z_1,\ldots,z_m)=\phi_{\tilde{L}}(\max_iz_i)$. Then $\psi(z_1,\ldots,z_m)$ is a bounded and continuous function. By Eq.~\eqref{eqn:limTpsi},
\begin{equation*}
\lim_{T\to\infty}\E\left[\phi_{\tilde{L}}\left(\max_{i=1,\ldots,m}\tilde{Z}_{T,i}\right)\right]  = \E\left[\phi_{\tilde{L}}\left(\max_{i=1,\ldots,m}G_i\right)\right].
\end{equation*}
For any $\tilde{L}>0$,
\begin{equation*}
\begin{aligned}
\E\left[\max_{i=1,\ldots,m}\tilde{Z}_{T,i}\right] &\geq \E\left[\phi_{\tilde{L}}\left(\max_{i=1,\ldots,m}\tilde{Z}_{T,i}\right)\right] + \E\left[\left(\tilde{L}+\max_{i=1,\ldots,m}\tilde{Z}_{T,i}\right)\mathbf 1_{\{\max_{i=1,\ldots,m}\tilde{Z}_{T,i}<-\tilde{L}\}}\right].
\end{aligned}
\end{equation*}
Moreover,
\allowdisplaybreaks[4]
\begin{equation*}
\begin{aligned}
&\E\left[\left(\tilde{L}+\max_{i=1,\ldots,m}\tilde{Z}_{T,i}\right)\mathbf 1_{\{\max_{i=1,\ldots,m}\tilde{Z}_{T,i}<-\tilde{L}\}}\right]\\
&\geq -\E\left[\left(\left|\max_{i=1,\ldots,m}\tilde{Z}_{T,i}\right|-\tilde{L}\right)\mathbf 1_{\{\max_{i=1,\ldots,m}\tilde{Z}_{T,i}<-\tilde{L}\}}\right]=  -\int_{\tilde{L}}^\infty\P\left[\left|\max_{i=1,\ldots,m}\tilde{Z}_{T,i}\right|>u\right]du\\
& \geq -\int_{\tilde{L}}^\infty m \max_{i=1,\ldots,m}\P\left[\left|\tilde{Z}_{T,i}\right|>u\right]du\geq -2m\int_{\tilde{L}}^\infty\left(1+\frac{1}{u^2}\right)\exp\left(-\frac{u^2}{2}\right)du\\
&=-\frac{2m}{\tilde{L}}\exp\left(-\frac{\tilde{L}^2}{2}\right).
\end{aligned}
\end{equation*}
Therefore, for any $\tilde{L}>0$,
\begin{equation*}
\underset{T\to\infty}{\lim\inf}~\E\left[\max_{i=1,\ldots,m}\tilde{Z}_{T,i}\right]\geq \E\left[\phi_L\left(\max_{i=1,\ldots,m}G_i\right)\right]-\frac{2m}{\tilde{L}}\exp\left(-\frac{\tilde{L}^2}{2}\right).
\end{equation*}
Letting $\tilde{L}\to\infty$. Then 
\begin{equation*}
\underset{T\to\infty}{\lim\inf}~\E\left[\max_{i=1,\ldots,m}\tilde{Z}_{T,i}\right]\geq \E\left[\max_{i=1,\ldots,m}G_i\right].
\end{equation*}
Similarly, it can be shown that
\begin{equation*}
\underset{T\to\infty}{\lim\sup}~\E\left[\max_{i=1,\ldots,m}\tilde{Z}_{T,i}\right]\leq \E\left[\max_{i=1,\ldots,m}G_i\right].
\end{equation*}
By Eq.~\eqref{eqn:defoftildez}, we have
\begin{equation}
\label{eqn:limtsumzit}
\lim_{T\to\infty}\frac{1}{\sqrt{T}}\E\left[\max_{i=1,\ldots,m}\sum_{t=1}^TZ_{i,t}\right]= \E\left[\max_{i=1,\ldots,m}G_i\right].
\end{equation}

Third, we note that  $G_1,\ldots, G_m$ are independent standard normal random variables. Then it is known that \citep[see, e.g.,][]{galambos1978},
\begin{equation}
\label{eqn:limmaxg}
\lim_{m\to\infty}\frac{1}{\sqrt{2\ln m}}\E\left[\max_{i=1,\ldots,m}G_i\right]=1.
\end{equation}
Combining Eqs.~\eqref{eqn:boundoninfsup}, \eqref{eqn:lowerbdonrt}, \eqref{eqn:limtsumzit}, and \eqref{eqn:limmaxg}, we complete the proof.
\end{proof}

\subsection{Proof of Theorem \ref{thm:nearoptimaldivergingT}}

\begin{proof}
Let $i_t^*$ be the index of the arm that has the smallest cumulative loss up to the first $t$ rounds, $i_t^*\equiv  \arg\min_{i\in\AA}\sum_{s=1}^t\hat{l}(i,y_s)$.
Consider the following,
\begin{equation}
\label{eqn:decompABC}
\begin{aligned}
& \frac{1}{\eta_t}\ln\frac{\exp\left(-\eta_t\cdot\sum_{s=1}^{t-1}\hat{l}(i^*_{t-1},y_s)\right)}{\sum_{j=1}^m\exp\left(-\eta_t\cdot\sum_{s=1}^{t-1}\hat{l}(j,y_s)\right)} - \frac{1}{\eta_{t+1}}\ln\frac{\exp\left(-\eta_{t+1}\cdot\sum_{s=1}^{t}\hat{l}(i^*_{t},y_s)\right)}{\sum_{j=1}^m\exp\left(-\eta_{t+1}\cdot\sum_{s=1}^{t}\hat{l}(j,y_s)\right)}\\
& := A+B+C,
\end{aligned}
\end{equation}
where
\begin{equation}
\label{def:abc}
\begin{aligned}
A & =  \left(\frac{1}{\eta_{t+1}}-\frac{1}{\eta_t}\right) \ln\frac{\sum_{j=1}^m\exp\left(-\eta_{t+1}\cdot\sum_{s=1}^{t}\hat{l}(j,y_s)\right)}{\exp\left(-\eta_{t+1}\cdot\sum_{s=1}^{t}\hat{l}(i^*_{t},y_s)\right)},\\ 
B & =  \frac{1}{\eta_t}\ln\frac{\sum_{i=1}^m\exp\left(-\eta_{t+1}\cdot d_i \right)}{\sum_{j=1}^m\exp\left(-\eta_{t}\cdot d_j\right)},\\
C & =  \frac{1}{\eta_t}\ln\frac{\sum_{j=1}^m\exp\left(-\eta_{t}\cdot d_j\right)}{\sum_{i=1}^m\exp\left(-\eta_{t}\cdot d_i'\right)},
\end{aligned}
\end{equation}
and  $d_i\equiv\sum_{s=1}^{t}\left(\hat{l}(i,y_s)-\hat{l}(i^*_{t},y_s)\right)$ and $d'_i\equiv\sum_{s=1}^{t-1}\left(\hat{l}(i,y_s)-\hat{l}(i^*_{t-1},y_s)\right)$.
We now bound three terms, $A, B$, and $C$ in Eq.~\eqref{def:abc}, separately. 

First, we consider the term $A$ in Eq.~\eqref{def:abc}. Since $\eta_t=\sqrt{(8\ln m)/t}$, we have  $\eta_{t+1}<\eta_t$, and 
\begin{equation*}
A \leq  \left(\frac{1}{\eta_{t+1}}-\frac{1}{\eta_t}\right) \ln m. 
\end{equation*}

Second, we consider the term $B$  in Eq.~\eqref{def:abc}. 
By the definition of $i_t^*$, we have $d_i\geq 0$.
Let the random variable $\mathbf{d}\in\{d_i:i=1,\ldots,m\}$, which has a distribution function $\mathbf P$ and a probability mass function given by 
\begin{equation*}
p(d_i)=\frac{\exp(-\eta_{t+1}\cdot d_i)}{\sum_{j=1}^m\exp(-\eta_{t+1}\cdot d_j)}.
\end{equation*} 
By Jensen's inequality, we have
\begin{equation}
\label{eqn:jensenlnexp}
\begin{aligned}
\ln\frac{\sum_{i=1}^m\exp\left(-\eta_{t+1}\cdot d_i\right)}{\sum_{j=1}^m\exp\left(-\eta_{t}\cdot d_j\right)}  = -\ln\E\left[\exp\left((\eta_{t+1}-\eta_t)\mathbf{d}\right)\right]\leq (\eta_t-\eta_{t+1})\E[\mathbf d].
\end{aligned}
\end{equation}
Define the entropy of $\mathbf d$ as $\mathbb H(\mathbf d) = -\sum_{i=1}^mp(d_i)\ln p(d_i)$. Let $\mathbf U$ be a random
variable that has a uniform distribution function $\mathbf Q$ over the set $\{d_i:i=1,\ldots,m\}$, where its probability mass function is denoted by $q$. Then the  Kullback-Leibler divergence of $\mathbf P$ and $\mathbf Q$ is
\begin{equation}
\label{eqn:DPQ}
\mathbb D(\mathbf P||\mathbf Q) = \sum_{i=1,\ldots,m:p(d_i)>0}p(d_i)\ln\frac{p(d_i)}{q(d_i)}=\ln m-\mathbb H(\mathbf d).
\end{equation}
On the other hand,  
\begin{equation}
\label{eqn:bdonkldiv}
\begin{aligned}
\mathbb D(\mathbf P||\mathbf Q) & =-\sum_{i=1,\ldots,m:p(d_i)>0}p(d_i)\ln\frac{q(d_i)}{p(d_i)} \\
&\geq -\sum_{i=1,\ldots,m:p(d_i)>0}p(d_i)\left(\frac{q(d_i)}{p(d_i)}-1\right)\geq 0,
\end{aligned}
\end{equation}
where the second step used $\ln x\leq x-1$ for any $x>0$.
Combining Eqs.~\eqref{eqn:DPQ} and \eqref{eqn:bdonkldiv}, we have that $\mathbb H(\mathbf d)\leq \ln m$.  Thus,
\begin{equation}
\label{eqn:bdlmetat1}
\begin{aligned}
\ln m & \geq \mathbb H(\mathbf d) \\
& = \sum_{i=1}^m\exp(-\eta_{t+1} d_i)\left(\eta_{t+1} d_i+\ln\sum_{j=1}^m\exp(-\eta_{t+1}d_j)\right)\frac{1}{\sum_{j=1}^m\exp(-\eta_{t+1} d_j)}\\
& = \eta_{t+1} \E[\mathbf d] +\ln\sum_{j=1}^m\exp(-\eta_{t+1} d_j)\\
& \geq \eta_{t+1} \E[\mathbf d],
\end{aligned}
\end{equation}
where the last step used $\sum_{j=1}^m\exp(-\eta_{t+1} d_j)\geq 1$. Hence Eq.~\eqref{eqn:bdlmetat1} implies $\E[\mathbf d]\leq (\ln m)/\eta_{t+1}$. Since $\eta_t>\eta_{t+1}$, together with Eqs.~\eqref{eqn:jensenlnexp} and \eqref{eqn:bdlmetat1}, we have
\begin{equation*}
\ln\frac{\sum_{i=1}^m\exp(-\eta_{t+1}\cdot d_i)}{\sum_{j=1}^m\exp(-\eta_t\cdot d_i)}\leq\frac{\eta_t-\eta_{t+1}}{\eta_{t+1}}\ln m.
\end{equation*}
Therefore,
\begin{equation*}
\begin{aligned}
B & \leq \left(\frac{1}{\eta_{t+1}}-\frac{1}{\eta_t}\right)\ln m.
\end{aligned}
\end{equation*}

Finally, we consider the term $C$  in Eq.~\eqref{def:abc}. Equivalently, we can write
\begin{equation*}
C  = \frac{1}{\eta_t}\ln\frac{\exp\left(-\eta_{t}\cdot\sum_{s=1}^{t-1}\hat{l}(i^*_{t-1},y_s)\right)}{\exp\left(-\eta_{t}\cdot\sum_{s=1}^{t}\hat{l}(i^*_{t},y_s)\right)} + \frac{1}{\eta_t}\ln\frac{\sum_{j=1}^m\exp\left(-\eta_{t}\cdot\sum_{s=1}^{t}\hat{l}(j,y_s)\right)}{\sum_{i=1}^m\exp\left(-\eta_{t}\cdot\sum_{s=1}^{t-1}\hat{l}(i,y_s)\right)}.
\end{equation*}
We study the two terms on the right-hand side separately. For the first term, we have
\begin{equation*}
\frac{1}{\eta_t}\ln\frac{\exp\left(-\eta_{t}\cdot\sum_{s=1}^{t-1}\hat{l}(i^*_{t-1},y_s)\right)}{\exp\left(-\eta_{t}\cdot\sum_{s=1}^{t}\hat{l}(i^*_{t},y_s)\right)} = \sum_{s=1}^{t}\hat{l}(i^*_{t},y_s) - \sum_{s=1}^{t-1}\hat{l}(i^*_{t-1},y_s).
\end{equation*}
For the second term, we can equivalently write it as
\begin{equation*}
 \frac{1}{\eta_t}\ln\sum_{j=1}^m\frac{\exp\left(-\eta_{t}\cdot\sum_{s=1}^{t-1}\hat{l}(j,y_s)\right)}{\sum_{i=1}^m\exp\left(-\eta_{t}\cdot\sum_{s=1}^{t-1}\hat{l}(i,y_s)\right)}\exp\left(-\eta_t\cdot\hat{l}(j,y_t)\right).
 \end{equation*}
Then by Hoeffding's inequality,  an upper bound for the second term is
\begin{equation*}
\begin{aligned}
 -\sum_{j=1}^{m}\frac{\exp\left(-\eta_{t}\cdot\sum_{s=1}^{t-1}\hat{l}(j,y_s)\right)}{\sum_{i=1}^m\exp\left(-\eta_{t}\cdot\sum_{s=1}^{t-1}\hat{l}(i,y_s)\right)}\hat{l}(j,y_t)+\frac{\eta_t}{8} = -\bar{l}(\bp_t,y_t)+\frac{\eta_t}{8},
\end{aligned}
\end{equation*}
where $\bar{l}(\bp_t,y_t)$ is defined in Eq.~\eqref{eqn:defofbarl}.
Finally, we plug the bounds of $A$, $B$, and $C$ to Eq.~\eqref{eqn:decompABC} and obtain that
\begin{equation*}
\begin{aligned}
\bar{l}(\bp_t,y_t)&\leq \sqrt{\frac{\ln m}{8t}}+\left[  \sum_{s=1}^{t}\hat{l}(i^*_{t},y_s) - \sum_{s=1}^{t-1}\hat{l}(i^*_{t-1},y_s) \right]+2\left(\frac{1}{\eta_{t+1}}-\frac{1}{\eta_t}\right)\ln m\\
&\quad +\frac{1}{\eta_{t+1}}\ln\frac{\exp\left(-\eta_{t+1}\cdot\sum_{s=1}^{t}\hat{l}(i^*_{t},y_s)\right)}{\sum_{j=1}^m\exp\left(-\eta_{t+1}\cdot\sum_{s=1}^{t}\hat{l}(j,y_s)\right)}\\
&\quad - \frac{1}{\eta_t}\ln\frac{\exp\left(-\eta_t\cdot\sum_{s=1}^{t-1}\hat{l}(i^*_{t-1},y_s)\right)}{\sum_{j=1}^m\exp\left(-\eta_t\cdot\sum_{s=1}^{t-1}\hat{l}(j,y_s)\right)}.
\end{aligned}
\end{equation*} 
We apply the above inequality to each $t=1,\ldots,T$ and sum up using $\sum_{t=1}^T1/\sqrt{t}\leq 2\sqrt{T}$, $\eta_t=\sqrt{(8\ln m)/t}$, and
\begin{equation*}
\begin{aligned}
&\sum_{t=1}^T\left[  \sum_{s=1}^{t}\hat{l}(i^*_{t},y_s) - \sum_{s=1}^{t-1}\hat{l}(i^*_{t-1},y_s) \right] = \min_{i\in\AA}\sum_{t=1}^Tl(i,y_t),\\
&\sum_{t=1}^T\left(\frac{1}{\eta_{t+1}}-\frac{1}{\eta_t}\right)=\sqrt{\frac{T+1}{8\ln m}}-\sqrt{\frac{1}{8\ln m}},\\
&\sum_{t=1}^T\left(\frac{1}{\eta_{t+1}}\ln\frac{\exp\left(-\eta_{t+1}\cdot\sum_{s=1}^{t}\hat{l}(i^*_{t},y_s)\right)}{\sum_{j=1}^m\exp\left(-\eta_{t+1}\cdot\sum_{s=1}^{t}\hat{l}(j,y_s)\right)}\right.\\
&\quad\quad\quad\quad\quad\quad \left.-\frac{1}{\eta_t}\ln\frac{\exp\left(-\eta_t\cdot\sum_{s=1}^{t-1}\hat{l}(i^*_{t-1},y_s)\right)}{\sum_{j=1}^m\exp\left(-\eta_t\cdot\sum_{s=1}^{t-1}\hat{l}(j,y_s)\right)}\right)\leq\sqrt{\frac{\ln m}{8}}.
\end{aligned}
\end{equation*}
Therefore, the $\bar{R}_T$ defined in Eq.~\eqref{eqn:defofbarRT} can be bounded by,
\begin{equation*}
\begin{aligned}
\bar{R}_T & \leq\sqrt{\frac{T}{2}\ln m} + \sqrt{\frac{T+1}{2}\ln m}-\sqrt{\frac{\ln m}{8}} \\
&\leq 2\sqrt{\frac{T}{2}\ln m}+\sqrt{\frac{\ln m}{8}}.
\end{aligned}
\end{equation*}
Taking expectations on both sides of this inequality and combining the result with Eq.~\eqref{eqn:bdonhalIp}, we complete the proof. 
\end{proof}

\end{document}